\documentclass[letterpaper,english,11pt,aps,letter,superscriptaddress,nofootinbib,pre]{revtex4}
\usepackage[T1]{fontenc}
\usepackage[latin9]{inputenc}
\usepackage{xcolor}
\usepackage{pdfcolmk}
\usepackage{babel}

\usepackage{amsmath}
\usepackage{graphicx}
\usepackage{amssymb}
\usepackage{esint}
\PassOptionsToPackage{normalem}{ulem}
\usepackage{ulem}
\usepackage[unicode=true,pdfusetitle,
 bookmarks=true,bookmarksnumbered=false,bookmarksopen=false,
 breaklinks=false,pdfborder={0 0 1},backref=false,colorlinks=false]
 {hyperref}

\makeatletter

\pdfpageheight\paperheight
\pdfpagewidth\paperwidth

\providecommand{\tabularnewline}{\\}
\providecolor{lyxadded}{rgb}{1,0,0}
\providecolor{lyxdeleted}{rgb}{0,0,1}

\@ifundefined{textcolor}{}
{%
 \definecolor{BLACK}{gray}{0}
 \definecolor{WHITE}{gray}{1}
 \definecolor{RED}{rgb}{1,0,0}
 \definecolor{GREEN}{rgb}{0,1,0}
 \definecolor{BLUE}{rgb}{0,0,1}
 \definecolor{CYAN}{cmyk}{1,0,0,0}
 \definecolor{MAGENTA}{cmyk}{0,1,0,0}
 \definecolor{YELLOW}{cmyk}{0,0,1,0}
 }

\usepackage{ae,aecompl}

\makeatother

\begin{document}

\title{Staggered Schemes for Fluctuating Hydrodynamics}

\author{Florencio {}``Balboa'' Usabiaga}

\affiliation{Departamento de Física Teórica de la Materia Condensada, Univeridad
Autónoma de Madrid, Madrid 28049, Spain}

\author{John B. Bell}

\affiliation{Center for Computational Science and Engineering, Lawrence Berkeley
National Laboratory, Berkeley, CA, 94720}

\author{Rafael Delgado-Buscalioni}

\affiliation{Departamento de Física Teórica de la Materia Condensada, Univeridad
Autónoma de Madrid, Madrid 28049, Spain}

\author{Aleksandar Donev}

\email{donev@courant.nyu.edu}

\affiliation{Courant Institute of Mathematical Sciences, New York University,
New York, NY 10012}

\author{Thomas G. Fai}

\affiliation{Courant Institute of Mathematical Sciences, New York University,
New York, NY 10012}

\author{Boyce E. Griffith}

\affiliation{Leon H. Charney Division of Cardiology, Department of Medicine, New
York University School of Medicine, New York, NY 10016}

\author{Charles S. Peskin}

\affiliation{Courant Institute of Mathematical Sciences, New York University,
New York, NY 10012}
\begin{abstract}
We develop numerical schemes for solving the isothermal compressible
and incompressible equations of fluctuating hydrodynamics on a grid
with staggered momenta. We develop a second-order accurate spatial
discretization of the diffusive, advective and stochastic fluxes that
satisfies a discrete fluctuation-dissipation balance, and construct
temporal discretizations that are at least second-order accurate in
time deterministically and in a weak sense. Specifically, the methods
reproduce the correct equilibrium covariances of the fluctuating fields
to third (compressible) and second (incompressible) order in the time
step, as we verify numerically. We apply our techniques to model recent
experimental measurements of giant fluctuations in diffusively mixing
fluids in a micro-gravity environment {[}\emph{A. Vailati et. al.,
Nature Communications 2:290, 2011}{]}. Numerical results for the static
spectrum of non-equilibrium concentration fluctuations are in excellent
agreement between the compressible and incompressible simulations,
and in good agreement with experimental results for all measured wavenumbers.
\end{abstract}
\maketitle
\global\long\def\V#1{\boldsymbol{#1}}
\global\long\def\M#1{\boldsymbol{#1}}
\global\long\def\Set#1{\mathbb{#1}}

\global\long\def\D#1{\Delta#1}
\global\long\def\d#1{\delta#1}

\global\long\def\norm#1{\left\Vert #1\right\Vert }
\global\long\def\abs#1{\left|#1\right|}

\global\long\def\grad{\boldsymbol{\nabla}}

\global\long\def\avv#1{\langle#1\rangle}
\global\long\def\av#1{\left\langle #1\right\rangle }

\providecommand{\deleted}[1]{}

\section{Introduction}

At a molecular scale, fluids are not deterministic; the state of the
fluid is constantly changing and stochastic, even at thermodynamic
equilibrium. Stochastic effects are important for flows in new microfluidic,
nanofluidic and microelectromechanical devices \cite{Nanofluidics_Review};
novel materials such as nanofluids \cite{Nanofluids_Review}; biological
systems such as lipid membranes \cite{SIBM_Biomembrane}, Brownian
molecular motors \cite{BrownainMotor_Peskin}, nanopores \cite{BrownianBiology};
as well as processes where the effect of fluctuations is amplified
by strong non-equilibrium effects, such as combustion of lean flames,
capillary dynamics \cite{CapillaryNanowaves,StagerredFluct_Inhomogeneous},
and hydrodynamic instabilities \cite{BreakupNanojets,DropletSpreading,DropFormationFluctuations},
and others. Because they span the whole range of scales from the microscopic
to the macroscopic \cite{DiffusionRenormalization_PRL,FractalDiffusion_Microgravity},
fluctuations need to be consistently included in all levels of description
\cite{DSMC_Hybrid}. Thermal fluctuations are included in the fluctuating
Navier-Stokes (NS) equations and related continuum Langevin models
\cite{GardinerBook,vanKampen:07} through stochastic forcing terms,
as first proposed by Landau and Lifshitz \cite{Landau:Fluid}. Numerically
solving the continuum equations of \emph{fluctuating hydrodynamics}
\cite{FluctHydroNonEq_Book} is difficult because of the presence
of non-trivial dynamics at all scales and the existence of a nontrivial
invariant measure (equilibrium distribution).

Several numerical approaches for fluctuating hydrodynamics have been
proposed. The earliest work by Garcia \emph{et al}. \cite{Garcia:87}
developed a simple scheme for the stochastic heat equation and the
linearized one-dimensional fluctuating NS equations. Ladd and others
have included stress fluctuations in (isothermal) Lattice Boltzmann
methods for some time \cite{LB_SoftMatter_Review}. Moseler and Landman
\cite{BreakupNanojets} included the stochastic stress tensor of Landau
and Lifshitz in the lubrication equations and obtain good agreement
with their molecular dynamics simulation in modeling the breakup of
nanojets. Sharma and Patankar \cite{FluctuatingHydro_FluidOnly} developed
a fluid-structure coupling between a fluctuating incompressible solver
and suspended Brownian particles. Coveney, De Fabritiis, Delgado-Buscalioni
and co-workers have also used the fluctuating isothermal NS equations
in a hybrid scheme, coupling a continuum fluctuating solver to a molecular
dynamics simulation of a liquid \cite{FluctuatingHydro_Coveney,FluctuatingHydroMD_Coveney,FluctuatingHydroHybrid_MD}.
Atzberger, Kramer and Peskin have developed a version of the immersed
boundary method that includes fluctuations \cite{AtzbergerETAL:2007,SELM}.
Voulgarakis and Chu \cite{StagerredFluctHydro} developed a staggered
scheme for the isothermal compressible equations as part of a multiscale
method for biological applications, and a similar staggered scheme
was also described in Ref. \cite{Delgado:08}.

Some of us have recently developed techniques for analyzing the weak
accuracy of finite-volume methods for solving the types of stochastic
partial differential equations that appear in fluctuating hydrodynamics
\cite{LLNS_S_k}. The analysis emphasizes the necessity to maintain
fluctuation-dissipation balance in spatio-temporal discretizations
\cite{LLNS_S_k}, thus reproducing the Gibbs-Boltzmann distribution
dictated by equilibrium statistical mechanics. Based on previous work
by Bell \emph{et al.} \cite{Bell:07,Bell:09}, a \emph{collocated}
spatial discretization for the compressible equations of fluctuating
hydrodynamics has been developed and combined with a stochastic third-order
Runge-Kutta (RK3) temporal integrator \cite{LLNS_S_k}. The collocated
spatial discretization has been used to construct a strictly conservative
particle-continuum hybrid method \cite{DSMC_Hybrid} and to study
the contribution of advection by thermal velocities to diffusive transport
\cite{DiffusionRenormalization}.

A \emph{staggered} spatial discretization is advantageous for incompressible
flows because it leads to a robust idempotent discrete projection
operator \cite{CFD_Patankar,IBM_Staggered}. Staggered schemes have
previously been developed for isothermal compressible \cite{StagerredFluctHydro}
and incompressible flow \cite{FluctuatingHydro_FluidOnly}, without,
however, carefully assessing discrete fluctuation-dissipation balance.
Here we present and test an explicit compressible and a semi-implicit
incompressible scheme for fluctuating hydrodynamics on uniform staggered
grids. Both methods use closely-related spatial discretizations, but
very different temporal discretizations.\emph{ }In the spatial discretization,
we ensure an accurate spectrum of the steady-state fluctuations by
combining a locally-conservative finite-volume formulation, a non-dissipative
(skew-symmetric) advection discretization, and discretely dual divergence
and gradient operators. For compressible flow, we employ an explicit
RK3 scheme \cite{LLNS_S_k} since the time step is limited by the
speed of sound and the dissipative terms can be treated explicitly.
For incompressible flow, we use a semi-implicit unsplit method first
proposed in Ref. \cite{NonProjection_Griffith}, which allows us to
take large time steps that under-resolve the fast momentum diffusion
at grid scales but still obtain the correct steady-state covariances
of fluctuations.

\begin{figure}[h]
\begin{centering}
\includegraphics[width=0.99\textwidth]{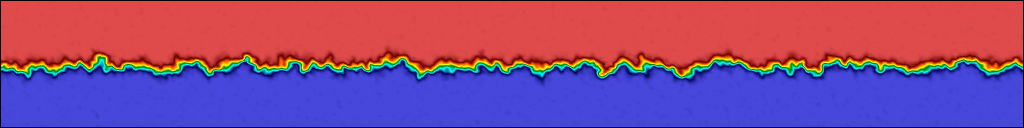}
\par\end{centering}

\begin{centering}
\includegraphics[width=0.99\textwidth]{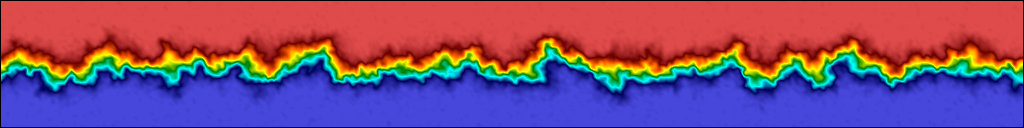}
\par\end{centering}

\begin{centering}
\includegraphics[width=0.99\textwidth]{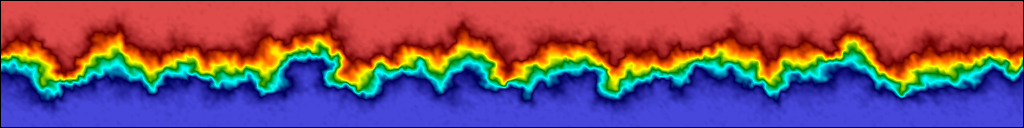}
\par\end{centering}

\caption{\label{fig:GiantFluctMixing}Snapshots of concentration showing the
development of a \emph{rough} diffusive interface between two miscible
fluids in zero gravity. We show three points in time (top to bottom),
starting from an initially perfectly flat interface (phase separated
system). These figures were obtained using the incompressible code
described in Section \ref{sub:Incompressible-Solver}.}
\end{figure}

Thermal fluctuations in non-equilibrium systems in which a constant
(temperature, concentration, velocity) gradient is imposed externally
exhibit remarkable behavior compared to equilibrium systems. Most
notably, external gradients can lead to \emph{enhancement} of thermal
fluctuations and to \emph{long-range} correlations between fluctuations
\cite{DSMC_Fluctuations_Shear,Mareschal:92,LongRangeCorrelations_MD,Zarate:04,FluctHydroNonEq_Book}.
This phenomenon can be illustrated by considering concentration fluctuations
in an isothermal mixture of two miscible fluids in the presence of
a strong concentration gradient $\grad c$, as in the early stages
of diffusive mixing between initially separated fluid components.
As illustrated in Fig. \ref{fig:GiantFluctMixing}, the interface
between the fluids, instead of remaining flat, develops large-scale
roughness that reaches a pronounced maximum until gravity or boundary
effects intervene. These \emph{giant fluctuations }\cite{GiantFluctuations_Theory,TemperatureGradient_Cannell,GiantFluctuations_ThinFilms}
during free diffusive mixing have been observed using light scattering
and shadowgraphy techniques \cite{GiantFluctuations_Nature,GiantFluctuations_Universal,GiantFluctuations_Microgravity,GiantFluctuations_Cannell,FractalDiffusion_Microgravity},
finding good but imperfect agreement between the predictions of a
simplified fluctuating hydrodynamic theory and experiments. Recent
experiments have taken advantage of the enhancement of the nonequilibrium
fluctuations in a microgravity environment aboard the FOTON M3 spaceship
\cite{GiantFluctuations_Microgravity,FractalDiffusion_Microgravity},
and demonstrated the appearance of fractal diffusive fronts like those
illustrated in Fig. \ref{fig:GiantFluctMixing}. In the absence of
gravity, the density mismatch between the two fluids does not change
the qualitative nature of the non-equilibrium fluctuations, and in
this work we focus on mixtures of dynamically-identical fluids.

Before discussing spatio-temporal discretizations, we review the continuum
formulation of the equations of fluctuating hydrodynamics and their
crucial properties in Section \ref{sec:Fluctuating-Hydrodynamics}.
In particular, we discuss the steady-state covariances of the fluctuating
fields for systems in thermal equilibrium as well as fluid mixtures
with an imposed concentration gradient. In Section \ref{sub:TemporalDiscretization}
we focus on the temporal discretization in the spirit of the method
of lines. For the compressible equations, we employ a previously-developed
explicit three-stage Runge-Kutta scheme that is third order weakly
accurate \cite{LLNS_S_k}. For the incompressible equations we employ
a second-order accurate predictor-corrector approach, each stage of
which is a semi-implicit (Crank-Nicolson) discretization of the Stokes
equations, solved effectively using a projection method as a preconditioner
\cite{NonProjection_Griffith}. In Section \ref{sub:SummarySpatial}
we describe a conservative staggered spatial discretization of the
diffusive, stochastic and advective fluxes. We maintain discrete fluctuation-dissipation
balance \cite{LLNS_S_k,DFDB} by ensuring duality between the discrete
divergence and gradient operators, and by using a skew-adjoint discretization
of advection. We verify the weak order of accuracy for both the compressible
and incompressible algorithms in Section \ref{sec:NumericalTests}.
In Section \ref{sec:Giant-Fluctuations} we model the non-equilibrium
concentration fluctuations in a fluid mixture under an applied temperature
gradient, and compare the numerical results to recent experimental
measurements \cite{GiantFluctuations_Microgravity,FractalDiffusion_Microgravity}.

\subsection{\label{sub:DetEquations}Deterministic Hydrodynamic Equations}

Motivated by the microgavity experiments studying giant fluctuations
\cite{GiantFluctuations_Microgravity,FractalDiffusion_Microgravity},
we consider an ideal solution of a macromolecule with molecular mass
$M$ in a solvent. At the macroscopic level, the hydrodynamics of
such a mixture can be modeled with an extended set of Navier-Stokes
equations for the mass density $\rho=\rho_{1}+\rho_{2}$, where $\rho_{1}$
is the mass density of the solute, $\V v$ is the center-of-mass velocity,
$c=\rho_{1}/\rho$ is the mass concentration, and $T$ is the temperature
\cite{FluctHydroNonEq_Book,Bell:09}. In many situations of interest
the temperature $T\left(\V r,t\right)\equiv T\left(\V r\right)$ can
be taken as fixed \cite{FluctHydroNonEq_Book,GiantFluctuations_Cannell,GiantFluctuations_Theory}
since temperature fluctuations do not couple to other variables. The
fixed-temperature \emph{compressible} NS equations for an ideal mixture
of two miscible fluids are \begin{align}
D_{t}\rho= & -\rho\left(\grad\cdot\V v\right)\label{eq:LLNS_rho}\\
\rho\left(D_{t}\V v\right)= & -\grad P+\grad\cdot\left[\eta\overline{\grad}\V v+\zeta\left(\grad\cdot\V v\right)\M I\right]+\V f_{v}\label{eq:LLNS_comp_v}\\
\rho\left(D_{t}c\right)= & \grad\cdot\left[\rho\chi\left(\grad c+c\left(1-c\right)S_{T}\grad T\right)\right]+f_{c},\label{eq:LLNS_comp_c}\end{align}
supplemented with appropriate boundary conditions. Here $D_{t}\square=\partial_{t}\square+\V v\cdot\grad\left(\square\right)$
is the advective derivative, $\overline{\grad}\V v=(\grad\V v+\grad\V v^{T})-2\left(\grad\cdot\V v\right)\M I/3$
is the symmetrized strain rate, $P\left(\rho,c;T\right)$ is the pressure
as given by the equation of state, and $\V f_{v}$ and $f_{c}$ are
external forcing (source) terms. The shear viscosity $\eta$, bulk
viscosity $\zeta$, mass diffusion coefficient $\chi$, and Soret
coefficient $S_{T}$, can, in general, depend on the state. We make
several physically-motivated approximations, including neglecting
barodiffusion, as we describe and justify next.

We will assume that the two species in the mixture are almost identical,
meaning that none of the fluid properties are affected by concentration.
In this sense the macromolecules are assumed to be simple (passive)
tracer particles. This is a reasonable approximation for small concentrations
$c\ll1$, since the presence of small amounts of macromolecule causes
small changes in the properties of the solution. In the giant fluctuation
experiments in microgravity conditions the concentration is at most
a few percent \cite{GiantFluctuations_Microgravity,FractalDiffusion_Microgravity},
justifying the assumption that the equation of state is independent
of concentration, $P\left(\rho,c;T\right)=P\left(\rho;T\right)$.
This approximation also allows us to neglect barodiffusion since the
barodiffusion coefficient is a thermodynamic rather than a transport
coefficient and vanishes for such an equation of state.

Because the temperature varies by only a few percent across the sample
in the giant fluctuation experiments modeled in Section \ref{sec:Giant-Fluctuations},
we take the system to be \emph{isothermal} and thus the temperature
$T=T_{0}$ is constant. However, we retain the crucial Soret term
by taking $S_{T}\grad T$ to be a specified constant. Note that the
Soret term is a transport coefficient unlike the barodiffusion coefficient
and can be positive or negative.

For liquids, the equation of state is usually very stiff, which means
that the (isothermal) sound speed $c_{T}^{2}=\partial P/\partial\rho$
is very large. Density is therefore nearly constant, and an incompressible
approximation will be appropriate as a means to avoid the stiffness.
An alternative, employed for example in the Lattice-Boltzmann method,
is to keep the simpler compressible equations (and thus avoid elliptic
constraints), but make the speed of sound much smaller than the actual
speed of sound, but still large enough that density variations are
negligible. This is the sense in which we will use the compressible
equations (\ref{eq:LLNS_rho},\ref{eq:LLNS_comp_v},\ref{eq:LLNS_comp_c}),
although we emphasize that there are situations in which it is actually
important to solve these equations with the proper equation of state
\cite{StagerredFluctHydro,DirectForcing_Balboa}.

Under the assumption that density variations are small, it is not
important what precise dependence of the transport coefficients and
equation of state on the density is used. We will therefore assume
that $P=P(\rho)=P_{0}+\left(\rho-\rho_{0}\right)c_{T}^{2}$, where
$c_{T}$ is a spatially-constant isothermal speed of sound. The value
of $c_{T}$ can be a parameter that lets us tune the compressibility
or the physical speed of sound. Furthermore, we will assume that the
viscosity and Soret coefficient are constants independent of the density,
and that the product $\rho\chi=\rho_{0}\chi_{0}$ is constant. Recalling
that in the experiments $c\ll1$ so that $c\left(1-c\right)\approx c$,
all of these approximations allows us to write the viscous term in
the momentum equation in the {}``Laplacian'' form\begin{equation}
\grad\cdot\left[\eta\overline{\grad}\V v+\zeta\left(\grad\cdot\V v\right)\M I\right]\rightarrow\eta\grad^{2}\V v+\left(\zeta+\frac{\eta}{3}\right)\grad\left(\grad\cdot\V v\right).\label{eq:visc_Laplacian}\end{equation}
Similarly, the diffusive term in the concentration equation can be
written as \begin{equation}
\grad\cdot\left[\rho\chi\left(\grad c+c\left(1-c\right)S_{T}\grad T\right)\right]\rightarrow\rho\left[\chi\grad^{2}c+\grad\cdot\left(c\V v_{s}\right)\right],\label{eq:diff_Laplacian}\end{equation}
where the spatially-constant velocity difference between the two species
is denoted with $\V v_{s}=\chi S_{T}\grad T$.

With all of these simplifications, the equations we actually solve
numerically are\begin{align}
D_{t}\rho= & -\rho\left(\grad\cdot\V v\right)\label{eq:LLNS_rho_simp}\\
\rho\left(D_{t}\V v\right)= & -c_{T}^{2}\grad\rho+\eta\grad^{2}\V v+\left(\zeta+\frac{\eta}{3}\right)\grad\left(\grad\cdot\V v\right)+\V f_{v}\label{eq:LLNS_comp_v_simp}\\
\rho\left(D_{t}c\right)= & \rho\left[\chi\grad^{2}c+\grad\cdot\left(c\V v_{s}\right)\right]+f_{c},\label{eq:LLNS_comp_c_simp}\end{align}
where all model parameters are constants.

We note that none of the simplifying approximations we make above
are necessary in principle. At the same time, not making such approximations
requires knowing a number of physical properties of the fluids, for
example, the concentration dependence of the Soret coefficient $S_{T}$.
Such information is difficult to obtain experimentally, and in any
case, the known dependence is very weak and we believe it will not
affect the results we present to within measurement or statistical
error bars. Furthermore, accounting for the concentration dependence
of the equation of state in the incompressible limit requires using
variable-density low Mach number equations \cite{LowMachMinicourse,LowMachAcoustics}
instead of the incompressible equations, since in general $\grad\cdot\V v\neq0$
\cite{Cahn-Hilliard_QuasiIncomp}. Extension of our algorithms to
these variable-density variable-coefficient low Mach equations is
possible but nontrivial, and will be considered in future work.

\section{\label{sec:Fluctuating-Hydrodynamics}Fluctuating Hydrodynamics}

At mesoscopic scales the hydrodynamic behavior of fluids can be described
with continuum stochastic PDEs of the Langevin type \cite{GardinerBook,vanKampen:07},
as proposed by Landau and Lifshitz \cite{Landau:Fluid} and later
justified by formal coarse-graining procedures \cite{LLNS_Espanol}.
Such equations can formally be justified as a central limit theorem
for the Gaussian behavior of the thermal fluctuations around the mean,
at least in certain simpler systems \cite{Lebowitz:88,Bertini:97}.
What emerges is that the (mesoscopic) thermal fluctuations can be
described by the very same hydrodynamic equations describing the macroscopic
behavior, \emph{linearized} around the mean solution, and with added
stochastic forcing terms that ensure a fluctuation-dissipation balance
principle \cite{FluctuationDissipation_Kubo}. Solving these linearized
equations numerically requires first solving the deterministic equations
for the mean, and then solving the fluctuating equations linearized
around the mean. The linearization typically contains many more terms
than the nonlinear deterministic terms due to the chain rule. Such
a two step process is much more cumbersome then solving the nonlinear
equations. Furthermore, a linearization has no hope of capturing any
possible nonlinear feedback of the fluctuations on the mean flow,
which is known to have physical significance \cite{DiffusionRenormalization}.

Therefore, we follow an alternative approach in which the stochastic
forcing terms are directly added to the nonlinear equations (\ref{eq:LLNS_rho_simp},\ref{eq:LLNS_comp_v_simp},\ref{eq:LLNS_comp_c_simp}),
but with an amplitude proportional to a parameter $\epsilon$ that
controls how far from linearity the equations are. In fluctuating
hydrodynamics, to ensure mass and momentum conservation, the stochastic
terms are the divergence of a stochastic flux,\begin{equation}
\V f_{v}=\epsilon^{\frac{1}{2}}\grad\cdot\M{\Sigma},\mbox{ and }f_{c}=\epsilon^{\frac{1}{2}}\grad\cdot\M{\Psi},\label{eq:stoch_forcing}\end{equation}
where the capital Greek letters denote stochastic fluxes that are
modeled as white-noise Gaussian random fields. A detailed discussion
of why there are no diffusive and stochastic fluxes in the density
equation is given in Ref. \cite{BrennerModification_Ottinger}. For
the linearized equations, we can fix $\epsilon=1$, and the covariances
of $\M{\Sigma}$ and $\M{\Psi}$ can be derived from the fluctuation-dissipation
balance principle, as explained well in the book \cite{FluctHydroNonEq_Book}.
The covariance of the stochastic stress tensor $\M{\Sigma}$ is not
a positive definite matrix so there are many choices for how to express
the stochastic stress, especially if additional bulk viscosity is
included \cite{OttingerBook}. We have based our implementation on
a formulation that requires the fewest possible random numbers \cite{LLNS_Espanol,DiscreteLLNS_Espanol},
\begin{eqnarray}
\M{\Sigma}=\M{\Sigma}_{s}+\M{\Sigma}_{p} & = & \sqrt{2\eta k_{B}T}\,\widetilde{\M{\mathcal{W}}}_{\V v}+\left(\sqrt{\frac{\zeta k_{B}T}{3}}-\frac{\sqrt{2\eta k_{B}T}}{3}\right)\,\mbox{Tr}\left(\widetilde{\M{\mathcal{W}}}_{\V v}\right)\M I,\label{stoch_flux_covariance}\\
\M{\Psi} & = & \sqrt{2\chi\rho M\; c(1-c)}\,\M{\mathcal{W}}_{c}\label{eq:stoch_conc_flux}\end{eqnarray}
where $\widetilde{\M{\mathcal{W}}}_{\V v}=(\M{\mathcal{W}}_{\V v}+\M{\mathcal{W}}_{\V v}^{T})/\sqrt{2}$
is a symmetric Gaussian random tensor field, and the $\sqrt{2}$ in
the denominator accounts for the reduction in variance due to the
averaging. Here $\M{\mathcal{W}}_{\V v}$ and $\M{\mathcal{W}}_{c}$
are mutually-uncorrelated white-noise random Gaussian tensor and vector
fields with uncorrelated components, \begin{align}
\av{\mathcal{W}_{ij}^{(\V v)}(\V r,t)\mathcal{W}_{kl}^{(\V v)}(\V r^{\prime},t^{\prime})} & =\left(\delta_{ik}\delta_{jl}\right)\delta(t-t^{\prime})\delta(\V r-\V r^{\prime})\label{eq:W_v_cov}\\
\av{\mathcal{W}_{i}^{(c)}(\V r,t)\mathcal{W}_{j}^{(c)}(\V r^{\prime},t^{\prime})} & =\left(\delta_{ij}\right)\delta(t-t^{\prime})\delta(\V r-\V r^{\prime}).\label{eq:W_c_cov}\end{align}
Similar covariance expressions apply in the Fourier domain as well
if position $\V r$ (time $t$) is replaced by wavevector $\V k$
(wavefrequency $\omega$), and $\av{\mathcal{W}_{\alpha}\mathcal{W}_{\beta}}$
is replaced by $\av{\widehat{\mathcal{W}}_{\alpha}\widehat{\mathcal{W}}_{\beta}^{\star}}$,
where star denotes complex conjugate (more generally, we denote an
adjoint of a matrix or linear operator with a star). We recall that
we take the temperature $T$ to be spatially-constant.

It is important to emphasize here that the non-linear fluctuating
NS equations, with the white-noise stochastic forcing terms (\ref{eq:stoch_forcing}),
are ill-defined because the solution should be a distribution rather
than a function and the nonlinear terms cannot be interpreted in the
sense of distributions. The nonlinear equations can be interpreted
using a small-scale \emph{regularization} (smoothing) of the stochastic
forcing, along with a suitable renormalization of the transport coefficients
\cite{DiffusionRenormalization_I,DiffusionRenormalization_PRL}. Such
a regularization is naturally provided by the discretization or coarse-graining
\cite{DiscreteLLNS_Espanol} length scale. As long as there are sufficiently
many molecules per hydrodynamic cell the fluctuations will be small
and the behavior of the nonlinear equations will closely follow that
of the \emph{linearized} equations of fluctuating hydrodynamics, which
can be given a precise meaning \cite{DaPratoBook}. This can be checked
by reducing $\epsilon$ to the point where the observed spatio-temporal
correlations of the fluctuations begin to scale linearly with $\epsilon$,
indicating nonlinear effects are negligible. In all of the simulations
reported here, we have used $\epsilon=1$ but have checked that using
a very small $\epsilon$ and then rescaling the covariance of the
fluctuations by $\epsilon^{-1}$ gives indistinguishable results to
within statistical errors.

Note that for the linearized equations the noise is additive since
the covariance of the stochastic forcing terms is to be evaluated
at the mean around which the linearization is performed. That is,
in the linearized equations ($\epsilon\rightarrow0$) one should read
(\ref{eq:stoch_conc_flux}) as $\M{\Psi}=\sqrt{2\chi\rho M\;\bar{c}(1-\bar{c})}\,\M{\mathcal{W}}_{c}$
where $\bar{c}$ is the solution of the deterministic equations. Therefore,
there is no Ito-Stratonovich difficulty in interpreting the stochastic
terms, and we use the (ambiguous) {}``Langevin'' notation that is
standard in the physics literature, instead of the differential notation
more common in the literature on stochastic differential equations.

It is important to observe that even when linearized, (\ref{eq:LLNS_rho_simp},\ref{eq:LLNS_comp_v_simp},\ref{eq:LLNS_comp_c_simp})
is a very challenging system of multiscale equations. Even a single
stochastic advection-diffusion equation such as (\ref{eq:LLNS_comp_v_simp})
is inherently multiscale because thermal fluctuations span the whole
range of spatio-temporal scales from the microscopic to the macroscopic,
specifically, \emph{all} modes of the spatial discretization have
a non-trivial stochastic dynamics that must be reproduced by the numerical
method. Including the density equation (\ref{eq:LLNS_rho_simp}) in
the system of equations leads to fast sound wave modes that make the
compressible equations stiff even in the deterministic setting. Finally,
in most applications of interest the concentration diffusion is much
slower than the momentum diffusion, leading to additional stiffness
and multiscale nature of the equations, as we discuss further in Section
\ref{sec:Giant-Fluctuations}.

\subsection{Incompressible Equations}

If density variations are negligible, $\rho=\rho_{0}=\mbox{const.}$,
we obtain the \emph{incompressible} approximation to the hydrodynamic
equations (\ref{eq:LLNS_rho_simp},\ref{eq:LLNS_comp_v_simp},\ref{eq:LLNS_comp_c_simp})
\cite{FluctHydroNonEq_Book},\begin{align}
\partial_{t}\V v= & -\grad\pi-\grad\cdot\left(\V v\V v^{T}\right)+\nu\grad^{2}\V v+\rho^{-1}\V f_{v}\label{eq:LLNS_incomp_v}\\
= & \M{\mathcal{P}}\left[-\V v\cdot\grad\V v+\nu\grad^{2}\V v+\rho^{-1}\V f_{v}\right]\nonumber \\
\partial_{t}c= & -\grad\cdot\left[c\left(\V v-\V v_{s}\right)\right]+\chi\grad^{2}c+\rho^{-1}f_{c},\label{eq:LLNS_incomp_c}\end{align}
where $\nu=\eta/\rho$, and $\V v\cdot\grad c=\grad\cdot\left(c\V v\right)$
and $\V v\cdot\grad\V v=\grad\cdot\left(\V v\V v^{T}\right)$ because
of incompressibility, $\grad\cdot\V v=0$. Here $\M{\mathcal{P}}$
is the orthogonal projection onto the space of divergence-free velocity
fields, $\M{\mathcal{P}}=\M I-\M{\mathcal{G}}\left(\M{\mathcal{D}}\M{\mathcal{G}}\right)^{-1}\M{\mathcal{D}}$
in real space, where $\M{\mathcal{D}}\square\equiv\grad\cdot\square$
denotes the divergence operator and $\M{\mathcal{G}}\equiv\grad$
the gradient operator. With periodic boundaries we can express all
operators in Fourier space and $\widehat{\M{\mathcal{P}}}=\M I-k^{-2}(\V k\V k^{\star})$
, where $\V k$ is the wavenumber.

Fluctuations can be included in the incompressible equations via the
stochastic forcing terms\begin{eqnarray*}
\rho^{-1}\V f_{v} & =\rho^{-1}\epsilon^{\frac{1}{2}}\grad\cdot\M{\Sigma}_{s}= & \grad\cdot\left(\sqrt{2\epsilon\nu\rho^{-1}k_{B}T}\,\widetilde{\M{\mathcal{W}}}_{\V v}\right)\\
\rho^{-1}f_{c} & =\rho^{-1}\epsilon^{\frac{1}{2}}\grad\cdot\M{\Psi}= & \grad\cdot\left[\sqrt{2\epsilon\chi\rho^{-1}M\, c(1-c)}\,\M{\mathcal{W}}_{c}\right]\end{eqnarray*}
Note that it is not necessary to include the stochastic pressure fluctuations
$\M{\Sigma}_{p}$ in (\ref{stoch_flux_covariance}) in the incompressible
velocity equation (\ref{eq:LLNS_incomp_v}) since the projection eliminates
any non-zero trace component of the stress tensor. In our formulation,
we use a strictly symmetric stochastic stress tensor $\M{\Sigma}_{s}$
in the incompressible equations. This is based on physical arguments
about local angular momentum conservation \cite{AngularMomentum_Hydrodynamics,AngularMomentumNanoflow}.
At the same time, the only thing that matters in the Fokker-Planck
description is the covariance of the stochastic forcing in the velocity
equation $\M{\mathcal{P}}\M D\M{\Sigma}_{s}$. This covariance is
determined from the fluctuation dissipation balance principle,\begin{equation}
\av{\left(\M{\mathcal{P}}\M{\mathcal{D}}\M{\Sigma}_{s}\right)\left(\M{\mathcal{P}}\M{\mathcal{D}}\M{\Sigma}_{s}\right)^{\star}}=\M{\mathcal{P}}\M{\mathcal{D}}\av{\M{\Sigma}_{s}\M{\Sigma}_{s}^{\star}}\M{\mathcal{D}}\M{\mathcal{P}}=\M{\mathcal{P}}\M{\mathcal{L}}\M{\mathcal{P}},\label{eq:FD_incomp}\end{equation}
where $\M{\mathcal{L}}$ is the vector Laplacian operator. Because
$\M{\mathcal{P}}$ and $\M{\mathcal{D}}$ have non-trivial null spaces,
(\ref{eq:FD_incomp}) does not uniquely determine the covariance of
the stochastic stress. In fact, it can easily be shown by going to
the Fourier domain that one can have a nonsymmetric component to the
stochastic stress without violating (\ref{eq:FD_incomp}). We believe
that the stress tensor should be symmetric since we do not include
an additional equation for the intrinsic spin (angular momentum) density.
This is appropriate for fluids composed of {}``point'' particles;
however, recent molecular dynamics simulations have shown that for
molecular liquids there can be non-trivial coupling between the linear
and spin momentum densities \cite{AngularMomentumNanoflow}. While
spin density has been included in the fluctuating hydrodynamics equations
\cite{AngularMomentum_Hydrodynamics} at the theoretical level, we
are not aware of any numerical simulations of such equations and do
not consider an angular momentum equation in this work.

\subsection{\label{sub:ContinuumSpectra}Steady-State Covariances}

The means and spatio-temporal covariances of the fluctuating fields
fully characterize the Gaussian solution of the linearized equations
\cite{LLNS_S_k}. Of particular importance is the steady-state covariance
of the fluctuating fields, which can be obtained for periodic systems
by linearizing the equations in the fluctuations and using a spatial
Fourier transform to decouple the different modes (wavevectors $\M k$).
This steady-state covariance in Fourier space is usually referred
to as a \emph{static structure factor} in the physical literature,
and represents the covariance matrix of the Fourier spectra of a typical
snapshot of the fluctuating fields. Note that it is in principle possible
to calculate the covariance of the fluctuations in non-periodic domains
as well \cite{RayleighBernard_LLNS}; however, these tedious calculations
offer little additional physical insight over the simple results presented
below. We will present numerical algorithms that can solve the fluctuating
equations with non-periodic of boundary conditions; however, periodic
conditions will be used to test the accuracy of the spatio-temporal
discretization by comparing to the simple theory.

At thermodynamic equilibrium, the fluctuations of the different hydrodynamic
variables are uncorrelated and white in space, that is, the equilibrium
variance is independent of the wavevector $\V k$ \cite{LLNS_S_k},
in agreement with equilibrium statistical mechanics \cite{Landau:StatPhys1,Landau:Fluid}.
Consider first the fluctuating isothermal compressible NS equations
(\ref{eq:LLNS_rho_simp},\ref{eq:LLNS_comp_v_simp},\ref{eq:LLNS_comp_c_simp})
linearized around a uniform steady-state, $\left(\rho,\V v,c\right)=\left(\rho_{0}+\d{\rho},\,\V v_{0}+\d{\V v},\, c_{0}+\d c\right)$,
$T=T_{0}$. Because of Galilean invariance, the advective terms $\V v_{0}\cdot\grad\left(\square\right)$
due to the presence of a background flow do not affect the equilibrium
covariances (structure factors), which are found to be \cite{FluctHydroNonEq_Book,LLNS_S_k}\begin{eqnarray}
S_{\rho,\rho}= & \av{\left(\widehat{\d{\rho}}\right)\left(\widehat{\d{\rho}}\right)^{\star}} & =\rho_{0}k_{B}T_{0}/c_{T}^{2}\nonumber \\
\M S_{\V v,\V v}= & \av{(\widehat{\delta\V v})(\widehat{\d{\V v}})^{\star}} & =\rho_{0}^{-1}k_{B}T_{0}\,\M I\nonumber \\
S_{c,c}= & \av{\left(\widehat{\d c}\right)\left(\widehat{\d c}\right)^{\star}} & =M\rho_{0}^{-1}\, c_{0}(1-c_{0}).\label{eq:S_equilibrium}\end{eqnarray}
At equilibrium, there are no cross-correlations between the different
variables, for example, $\M S_{c,\V v}=\av{(\widehat{\delta c})(\widehat{\d{\V v}})^{\star}}=\V 0$.
The equilibrium variance of the spatial average of a given variable
over a cell of volume $\D V$ can be obtained by dividing the corresponding
structure factor by $\D V$, for example, the variance of the concentration
is $\av{\left(\d{\rho}\right)^{2}}=\rho_{0}k_{B}T_{0}/\left(c_{T}^{2}\D V\right)$.
In the incompressible limit, $c_{T}\rightarrow\infty$, the density
fluctuations vanish and $\rho\approx\rho_{0}$.

Out of thermodynamic equilibrium, there may appear long-ranged correlations
between the different hydrodynamic variables \cite{FluctHydroNonEq_Book}.
As a prototypical example of such non-equilibrium fluctuations, we
focus on the incompressible equations (\ref{eq:LLNS_incomp_v},\ref{eq:LLNS_incomp_c})
in the presence of an imposed concentration gradient $\grad\bar{c}$.
The spatial non-uniformity of the mean concentration when there is
a gradient breaks the translational symmetry and the Fourier transform
no longer diagonalizes the equations. We focus our analysis and test
our numerical schemes on a periodic approximation in which we linearize
around a uniform background state $\left(\V v,c\right)=\left(\d{\V v},\, c_{0}+\d c\right)$,
as suggested and justified in the physics literature on long-range
nonequilibrium correlations \cite{FluctHydroNonEq_Book,Zarate:04,GiantFluctuations_Theory,GiantFluctuations_Cannell}.
In such a periodic approximation we cannot have a gradient in the
steady-state average concentration $c_{0}$ but we can mimic the effect
of the advective term $\V v\cdot\grad c_{0}$ with an additional term
$\V v\cdot\left(\grad\bar{c}\right)$ in the concentration equation.
This is justified if the concentration gradient is weak, and leads
to the linearized equations in a periodic domain,\begin{eqnarray}
\partial_{t}\left(\d{\V v}\right) & = & \M{\mathcal{P}}\left[\nu\grad^{2}\left(\d{\V v}\right)+\grad\cdot\left(\sqrt{2\nu\rho_{0}^{-1}k_{B}T_{0}}\,\M{\mathcal{W}}_{\V v}\right)\right]\nonumber \\
\partial_{t}\left(\d c\right) & = & -\left(\grad\bar{c}\right)\cdot\left(\d{\V v}\right)+\chi\grad^{2}\left(\d c\right)+\grad\cdot\left[\sqrt{2\chi\rho_{0}^{-1}M\, c_{0}(1-c_{0})}\,\M{\mathcal{W}}_{c}\right].\label{eq:incomp_quasi_periodic}\end{eqnarray}
In the Fourier domain (\ref{eq:incomp_quasi_periodic}) is a collection
of stochastic differential equations, one system of linear additive-noise
equations per wavevector $\V k$, written in differential notation
as\begin{eqnarray}
d\left(\widehat{\delta\V v}\right) & = & -\nu\, k^{2}\left(\widehat{\delta\V v}\right)dt+i\sqrt{2\nu\rho_{0}^{-1}k_{B}T_{0}}\,\widehat{\M{\mathcal{P}}}\V k\cdot\left(d\M{\mathcal{B}}_{\V v}^{\left(\V k\right)}\right)\nonumber \\
d\left(\widehat{\delta c}\right) & = & -\left(\grad\bar{c}\right)\cdot\left(\widehat{\delta\V v}\right)dt-\chi k^{2}\left(\widehat{\delta c}\right)dt+i\sqrt{2\chi\rho_{0}^{-1}M\, c_{0}(1-c_{0})}\,\V k\cdot\left(d\M{\mathcal{B}}_{c}^{\left(\V k\right)}\right),\label{eq:c_v_Fourier}\end{eqnarray}
where we used that $\widehat{\M{\mathcal{P}}}\left(\widehat{\delta\V v}\right)=\widehat{\delta\V v}$.
Here $\V{\mathcal{B}}_{\V v}(t)$ is a tensor, and $\V{\mathcal{B}}_{c}(t)$
is a vector, whose components are independent Wiener processes. Note
that the velocity equation is not affected by the concentration gradient.
Given the model equations (\ref{eq:c_v_Fourier}), the explicit solution
for the matrix of static structure factors (covariance matrix)\[
\M{\mathcal{S}}=\left[\begin{array}{cc}
\M S_{\V v,\V v} & S_{c,\V v}^{\star}\\
S_{c,\V v} & S_{c,c}\end{array}\right]\]
can be obtained as the solution of a linear system resulting from
the stationarity condition $d\M{\mathcal{S}}=\M 0$. For a derivation,
see Eq. (30) in \cite{LLNS_S_k} or Eq. (3.10) in \cite{AMR_ReactionDiffusion_Atzberger}
and also Eq. (\ref{eq:DFDB_model}); below we simply quote the results
of these straightforward calculations.

\subsubsection{\label{sub:TheoryIncomp}Incompressible Velocity Fluctuations}

By considering the stationarity condition $d\M S_{\V v,\V v}=0$ it
can easily be seen that the equilibrium covariance of the velocities
is proportional to the projection operator, \begin{equation}
\M S_{\V v,\V v}=\rho_{0}^{-1}k_{B}T_{0}\,\widehat{\M{\mathcal{P}}}=\rho_{0}^{-1}k_{B}T_{0}\,\left[\M I-k^{-2}(\V k\V k^{\star})\right],\label{S_k_incompressible}\end{equation}
independent of the concentration gradient. In particular, the amplitude
of the velocity fluctuations at each wavenumber is constant and reduced
by one in comparison to the compressible equations,\begin{equation}
\mbox{Tr}\,\M S_{\V v,\V v}=\left\langle (\widehat{\d{\V v}})^{\star}(\widehat{\d{\V v}})\right\rangle =\left(d-1\right)\rho_{0}^{-1}k_{B}T_{0},\label{Trace_S_continuum}\end{equation}
where $d$ is the spatial dimension. This is a reflection of the fact
that one degree of freedom (i.e., one $k_{B}T/2$) is subtracted from
the kinetic energy due to the incompressibility constraint, which
eliminates the sound mode. In Appendix \ref{sec:ProjectionAccuracy}
we obtain the generalization of (\ref{S_k_incompressible}) to non-periodic
systems,\begin{equation}
\left\langle \V v\V v^{\star}\right\rangle =\rho_{0}^{-1}k_{B}T_{0}\,\left(\D V^{-1}\M{\mathcal{P}}\right).\label{C_vv_incompressible}\end{equation}

An alternative way of expressing the result (\ref{C_vv_incompressible})
is that all divergence-free modes have the same spectral power at
equilibrium. That is, if the fluctuating velocities are expressed
in any orthonormal basis for the space of velocities that satisfy
$\grad\cdot\V v=0$, at equilibrium the resulting random coefficients
should be uncorrelated and have unit variance. This will be useful
in Section \ref{sub:Incompressible-Solver} for examining the weak
accuracy of the spatio-temporal discretizations of the incompressible
equations. For periodic boundary conditions, such an orthonormal basis
is simple to construct in the Fourier domain and a Fourier transform
can be used project the velocity field onto this basis. In particular,
for all wavevectors the projection of the velocity fluctuations onto
the \emph{longitudinal} mode\begin{equation}
\hat{\V v}^{(1)}=k^{-1}\left[k_{x},\, k_{y},\, k_{z}\right],\label{eq:incomp_basis_1}\end{equation}
where $k=\left(k_{x}^{2}+k_{y}^{2}+k_{z}^{2}\right)^{1/2}$, should
be identically zero,\[
\hat{v}_{1}=(\widehat{\d{\V v}})\cdot\hat{\V v}=\frac{k_{x}}{k}\widehat{\d v_{x}}+\frac{k_{y}}{k}\widehat{\d v_{y}}+\frac{k_{z}}{k}\widehat{\d v_{z}}=k^{-1}\left(\V k\cdot\V v\right)=0.\]
A basis for the incompressible periodic velocity fields can be constructed
from the two \emph{vortical} modes \begin{eqnarray}
\hat{\V v}^{(2)} & = & \left(k_{x}^{2}+k_{y}^{2}\right)^{-1/2}\left[-k_{y},\, k_{x},\,0\right],\label{eq:incomp_basis_2}\\
\hat{\V v}^{(3)} & = & k^{-1}\left(k_{x}^{2}+k_{y}^{2}\right)^{-1/2}\left[k_{x}k_{z},\, k_{y}k_{z},\,-\left(k_{x}^{2}+k_{y}^{2}\right)\right],\label{eq:incomp_basis_3}\end{eqnarray}
and the projection of the fluctuating velocities onto these modes
has the equilibrium covariance\begin{equation}
\av{\hat{v}_{2}\hat{v}_{2}^{\star}}=\av{\hat{v}_{3}\hat{v}_{3}^{\star}}=\rho_{0}^{-1}k_{B}T_{0},\mbox{ while }\av{\hat{v}_{2}\hat{v}_{3}^{\star}}=0.\label{eq:incomp_basis_dfdb}\end{equation}
In two dimensions only $\hat{\V v}^{(1)}$ and $\hat{\V v}^{(2)}$
are present, and $\hat{\V v}^{(2)}$ is the $z$ component of the
vorticity and spans the subspace of diverence-free velocities. The
fact that the $(d-1)$ vortical modes have equal power leads to the
velocity variance (\ref{Trace_S_continuum}).

\subsubsection{\label{sub:Nonequilibrium-Fluctuations}Nonequilibrium Fluctuations}

When a macroscopic concentration gradient is present, the velocity
fluctuations affect the concentration via the linearized advective
term $\left(\grad\bar{c}\right)\cdot\V v$. Solving (\ref{eq:c_v_Fourier})
shows an \emph{enhancement} of the concentration fluctuations \cite{ExtraDiffusion_Vailati}
proportional to the square of the applied gradient,\begin{equation}
S_{c,c}=M\rho_{0}^{-1}\, c_{0}(1-c_{0})+\frac{k_{B}T}{\rho\chi(\nu+\chi)k^{4}}\left(\sin^{2}\theta\right)\,\left(\nabla\bar{c}\right)^{2},\label{eq:S_cc_neq}\end{equation}
where $\theta$ is the angle between $\V k$ and $\grad\bar{c}$,
$\sin^{2}\theta=k_{\perp}^{2}/k^{2}$. Furthermore, there appear \emph{long-range
correlations} between the concentration fluctuations and the fluctuations
of velocity parallel to the concentration gradient, proportional to
the applied gradient \cite{ExtraDiffusion_Vailati,DiffusionRenormalization_PRL},
\begin{equation}
\mathcal{S}_{c,v_{\Vert}}=\av{(\widehat{\delta c})(\widehat{\d v}_{\Vert}^{\star})}=-\frac{k_{B}T}{\rho(\nu+\chi)k^{2}}\left(\sin^{2}\theta\right)\,\nabla\bar{c}.\label{eq:S_c_vy}\end{equation}
The power-law divergence for small $k$ indicates long-range correlations
between $\delta c$ and $\d{\V v}$ and is the cause of the giant
fluctuation phenomenon studied in Section \ref{sec:Giant-Fluctuations}.

\section{\label{sec:SpatioTemporal}Spatio-Temporal Discretization}

Designing temporal discretizations for fluid dynamics is challenging
even without including thermal fluctuations. When there is no stochastic
forcing, our schemes revert to standard second-order discretizations
and can be analyzed with existing numerical analysis techniques. Here
we tackle the additional goal of constructing discretizations that,
in a weak sense, accurately reproduce the statistics of the continuum
fluctuations for the linearized equations. Note that achieving second-order
weak accuracy is much simpler for linear additive-noise equations
since in the linear case the solution is fully characterized by the
means and the correlation functions (time-dependent covariances).
In fact, one can use any method that is second-order in time in the
deterministic setting and also reproduces the correct static (equal-time)
covariance to second order, as explained in more detail in Ref. \cite{LLNS_S_k}.
The deterministic order of accuracy can be analyzed using standard
techniques, and the accuracy of the static covariances can be analyzed
using the techniques described in Ref. \cite{LLNS_S_k}. We emphasize
that the temporal integrators are only higher-order accurate in a
weak sense for the linearized equations of fluctuating hydrodynamics
\cite{LLNS_S_k,DFDB}.

Thermal fluctuations are added to a deterministic scheme as an additional
forcing term that represents the temporal average of a stochastic
forcing term over the time interval $\D t$ and over the spatial cells
of volume $\D V$ \cite{LLNS_S_k}. Because $\M{\mathcal{W}}$ is
white in space and time, the averaging adds an additional prefactor
of $\left(\D V\,\D t\right)^{-1/2}$ in front of the stochastic forcing.
In the actual numerical schemes, a {}``realization'' of a white-noise
field $\M{\mathcal{W}}$ is represented by a collection $\M W$ of
normally-distributed random numbers with mean zero and covariance
given by (\ref{eq:W_c_cov}) or (\ref{eq:W_v_cov}), with the identification\[
\M{\mathcal{W}}\longleftrightarrow\left(\D V\,\D t\right)^{-1/2}\M W.\]
Specifically, the stochastic fluxes (\ref{stoch_flux_covariance})
are discretized as\begin{align}
\M{\Sigma}_{s}=\sqrt{\frac{2\eta k_{B}T}{\D V\,\D t}}\,\widetilde{\M W}_{\V v},\mbox{ and } & \M{\Psi}=\sqrt{\frac{2\chi\rho M\; c(1-c)}{\D V\,\D t}}\,\M W_{c}.\label{disc_stoch_flux_covariance}\end{align}

A realization of $\M W$ is sampled using a pseudo-random number stream.
The temporal discretization of the stochastic forcing corresponds
to the choice of how many realizations of $\M W$ are generated per
time step, and how each realization is associated to specific points
in time inside a time step (e.g., the beginning, mid-point, or end-point
of a time step). The spatial discretization corresponds to the choice
of how many normal variates to generate per spatial cell, and how
to associate them with elements of the spatial discretization (e.g.,
cell centers, nodes, faces, edges). Once these choices are made, it
is simple to add the stochastic forcing to an existing deterministic
algorithm or code, while still accounting for the fact that white-noise
is not like a classical smooth forcing and cannot be evaluated pointwise.

\subsection{\label{sub:TemporalDiscretization}Temporal Discretization}

As a first step in designing a spatio-temporal discretization for
the compressible and incompressible equations of fluctuating hydrodynamics,
we focus on the temporal discretization. We assume that the time step
is fixed at $\D t$. The time step index is denoted with a superscript,
for example, $\V c^{n}$ denotes concentration at time $n\D t$ and
$\V W^{n}$ denotes a realization of $\M W$ generated at time step
$n$. 

In the next section, we will describe our staggered spatial discretization
of the crucial differential operators, denoted here rather generically
with a letter symbol in order to distinguish them from the corresponding
continuum operators. Specifically, let $\M G$ be the gradient (scalar$\rightarrow$vector),
$\M D$ the divergence (vector$\rightarrow$scalar), and $\M L=\M D\M G$
the Laplacian (scalar$\rightarrow$scalar) operator. When the divergence
operator acts on a tensor field $\M F$ such as a stress tensor $\M{\sigma}$,
it is understood to act component-wise on the $x$, $y$ and $z$
components of the tensor. Similarly, the gradient and Laplacian act
component-wise on a vector. An important property of the discrete
operators that we require to hold is that the divergence operator
is the negative adjoint of the gradient, $\M D=-\M G^{\star}$. This
ensures that the scheme satisfies a discrete version of the continuous
property, \[
\int_{\Omega}w\left[\grad\cdot\V v\right]\, d\V r=-\int_{\Omega}\V v\cdot\grad w\, d\V r\mbox{ if }\V v\cdot\V n_{\partial\Omega}=0\mbox{ or }\V v\mbox{ is periodic}\]
for any scalar field $w(\V r)$.

We define the weak order of accuracy of a temporal discretization
in terms of the mismatch between the steady-state covariance of the
continuum and the discrete formulations. With periodic boundary conditions
this would be the mismatch between the Fourier spectrum of a typical
snapshot of the true solution and the steady-state discrete spectrum
of the numerical solution \cite{LLNS_S_k}. This mismatch is typically
of the form $O(\D t^{k})$ for some integer $k\geq1$, implying that
for sufficiently small time steps the discrete formulation reproduces
the steady-state covariance of the continuum formulation. Note that
for the linearized equations a certain order of deterministic temporal
accuracy, combined with equal or higher order of accuracy of the steady-state
covariances, implies the same order of accuracy for all temporal correlations.
A theoretical analysis of the weak accuracy of the temporal discretizations
used in this work can be performed using the tools described in Ref.
\cite{LLNS_S_k} with some straightforward extensions \cite{DFDB};
here we simply state the main results and verify the order of weak
accuracy numerically.

\subsubsection{\label{sub:CompressibleTemporal}Compressible Equations}

Denoting the fluctuating field with $\V Q=\left(\rho,\V v,c\right)$,
the fluctuating compressible NS equations (\ref{eq:LLNS_rho_simp},\ref{eq:LLNS_comp_v_simp},\ref{eq:LLNS_comp_c_simp})
can be written as a general stochastic conservation law,\begin{equation}
\partial_{t}\V Q=-\M D\left[\V F(\V Q;t)-\V Z(\bar{\V Q},\V W)\right],\label{LLNS_general}\end{equation}
where $\M D$ is the divergence operator (acting component-wise on
each flux), $\V F(\V Q;t)$ is the deterministic and $\V Z=\left[0,\,\M{\Sigma},\,\M{\Psi}\right]$
is the discretization of the stochastic flux (\ref{disc_stoch_flux_covariance}).
We recall that the stochastic forcing amplitude is written as multiplicative
in the state; however, in the linearized limit of weak fluctuations
the strength of the stochastic forcing only depends on the \emph{mean}
state, which is well-approximated by the instantaneous state, $\bar{\V Q}(t)\approx\M Q(t)$.
Following \cite{Bell:07}, we base our temporal discretization of
(\ref{LLNS_general}) on the (optimal) three-stage low-storage strong
stability preserving \cite{SSP_Review} (originally called total variation
diminishing \cite{Gottlieb:98}) Runge-Kutta (RK3) scheme of Gottlieb
and Shu, ensuring stability in the inviscid limit without requiring
slope-limiting. The stochastic terms are discretized using two random
fluxes per time step, as proposed in Ref. \cite{LLNS_S_k}. This discretization
achieves third-order weak accuracy \cite{DFDB} for linear additive-noise
equations, while only requiring the generation of two Gaussian random
fields per time step. 

For each stage of our third-order Runge-Kutta scheme, a conservative
increment is calculated as\[
\D{\V Q}(\V Q,\V W;t)=-\D t\,\M D\V F(\V Q;\, t)+\D t\,\M D\V Z(\V Q,\V W).\]
Each time step of the RK3 algorithm is composed of three stages, the
first one estimating $\V Q$ at time $t=(n+1)\D t$, the second at
$t=(n+\frac{1}{2})\D t$, and the final stage obtaining a third-order
accurate estimate at $t=(n+1)\D t$. Each stage consists of an Euler-Maruyama
step followed by a weighted averaging with the value from the previous
stage, \begin{align}
\widetilde{\V Q}^{n+1}= & \V Q^{n}+\D{\V Q}\left(\V Q^{n},\V W_{1}^{n}\,;\, n\D t\right)\nonumber \\
\widetilde{\V Q}^{n+\frac{1}{2}}= & \frac{3}{4}\V Q^{n}+\frac{1}{4}\left[\widetilde{\V Q}^{n+1}+\D{\V Q}\left(\widetilde{\V Q}^{n+1},\V W_{2}^{n}\,;\,(n+1)\D t\right)\right]\nonumber \\
\V Q^{n+1}= & \frac{1}{3}\V Q^{n}+\frac{2}{3}\left[\widetilde{\V Q}^{n+\frac{1}{2}}+\D{\V Q}\left(\widetilde{\V Q}^{n+\frac{1}{2}},\V W_{3}^{n}\,;\,(n+\frac{1}{2})\D t\right)\right],\label{RK3_explicit_stages}\end{align}
where the stochastic fluxes between different stages are related to
each other via \begin{align}
\V W_{1}^{n}= & \V W_{A}^{n}+w_{1}\V W_{B}^{n}\nonumber \\
\V W_{2}^{n}= & \V W_{A}^{n}+w_{2}\V W_{B}^{n}\nonumber \\
\V W_{3}^{n}= & \V W_{A}^{n}+w_{3}\V W_{B}^{n},\label{RK3_optimal}\end{align}
and $\V W_{A}^{n}$ and $\V W_{B}^{n}$ are two independent realizations
of $\M W$ that are generated independently at each RK3 step. In this
work we used the weights derived in Ref. \cite{LLNS_S_k} based on
a linearized analysis, $w_{1}=-\sqrt{3}$, $w_{2}=\sqrt{3}$ and $w_{3}=0$.
More recent analysis based on the work in \cite{RK3_WeakAdditive}
shows that second-order weak accuracy is achieved for additive-noise
\emph{nonlinear} stochastic differential equations using the weights
\cite{DFDB}\[
w_{1}=\frac{\left(2\,\sqrt{2}-\sqrt{3}\right)}{5},\quad w_{2}=\frac{\left(-4\,\sqrt{2}-3\,\sqrt{3}\right)}{5},\mbox{ and }w_{3}=\frac{\left(\sqrt{2}+2\,\sqrt{3}\right)}{10}.\]
For the types of problems studied here nonlinearities play a minimal
role and either choice of the weights is appropriate.

\subsubsection{\label{sub:IncompressibleTemporal}Incompressible Equations}

The spatially-discretized equations (\ref{eq:LLNS_incomp_v},\ref{eq:LLNS_incomp_c})
can be written in the form \begin{eqnarray*}
\partial_{t}\V v+\M G\pi & = & \M A_{\V v}(\V v,c)+\nu\M L\V v+\rho^{-1}\V f_{v},\\
\partial_{t}c & = & \M A_{c}(\V v,c)+\chi\M Lc+\rho^{-1}f_{c},\\
\mbox{s.t. }\M D\V v & = & \V 0,\end{eqnarray*}
where $\M A(\V v,c)$ represent the non-diffusive deterministic terms,
such as the advective and Soret forcing terms , as well as any additional
terms arising from gravity or other effects. Fluctuations are accounted
for via the stochastic forcing terms\[
\rho^{-1}\V f_{v}\left(\widetilde{\V W}_{\V v}\right)=\M D\left[\sqrt{\frac{2\epsilon\nu\rho^{-1}k_{B}T}{\D V\,\D t}}\,\widetilde{\V W}_{\V v}\right]\mbox{ and }\rho^{-1}f_{c}\left(c,\M W_{c}\right)=\M D\left[\sqrt{\frac{2\epsilon\chi\rho^{-1}M\; c(1-c)}{\D V\,\D t}}\,\M W_{c}\right].\]

We base our temporal discretization on the second-order semi-implicit
deterministic scheme of Griffith \cite{NonProjection_Griffith}, a
predictor-corrector method in which the predictor step combines the
Crank-Nicolson method for the diffusive terms with the Euler method
for the remaining terms,\begin{eqnarray}
\frac{\tilde{\V v}^{n+1}-\V v^{n}}{\D t}+\M G\tilde{\pi}^{n+\frac{1}{2}} & = & \M A_{\V v}(\V v^{n},c^{n})+\nu\M L\left(\frac{\tilde{\V v}^{n+1}+\V v^{n}}{2}\right)+\rho^{-1}\V f_{v}\left(\widetilde{\V W}_{\V v}^{n}\right),\nonumber \\
\frac{\tilde{c}^{n+1}-c^{n}}{\D t} & = & \M A_{c}(\V v^{n},c^{n})+\chi\M L\left(\frac{\tilde{c}^{n+1}+c^{n}}{2}\right)+\rho^{-1}f_{c}\left(c^{n},\M W_{c}^{n}\right)\nonumber \\
\mbox{s.t. }\M D\tilde{\V v}^{n+1} & = & \V 0.\label{eq:predictor}\end{eqnarray}
The corrector stage combines Crank-Nicolson for the diffusive terms
with an explicit second-order approximation for the remaining deterministic
terms,\begin{eqnarray}
\frac{\V v^{n+1}-\V v^{n}}{\D t}+\M G\pi^{n+\frac{1}{2}} & = & \M A_{\V v}^{n+\frac{1}{2}}+\nu\M L\left(\frac{\V v^{n+1}+\V v^{n}}{2}\right)+\rho^{-1}\V f_{v}^{n+\frac{1}{2}},\nonumber \\
\frac{c^{n+1}-c^{n}}{\D t} & = & \M A_{c}^{n+\frac{1}{2}}+\chi\M L\left(\frac{c^{n+1}+c^{n}}{2}\right)+\rho^{-1}f_{c}^{n+\frac{1}{2}}\nonumber \\
\mbox{s.t. }\M D\V v^{n+1} & = & \V 0.\label{eq:corrector}\end{eqnarray}
Unlike a fractional-step scheme that splits the velocity and pressure
updates \cite{bellColellaGlaz:1989,almgrenBellSzymczak:1996}, this
approach simultaneously solves for the velocity and pressure and avoids
the need to determine appropriate {}``intermediate'' boundary conditions.
Importantly, no spurious boundary modes \cite{ProjectionModes_III,GaugeIncompressible_E}
arise due to the implicit velocity treatment even in the presence
of physical boundaries, which is especially important for fluctuating
hydrodynamics since all of the modes are stochastically forced \cite{DFDB}.

The concentration equation in (\ref{eq:corrector}) {[}and similarly
in (\ref{eq:predictor}){]} is a linear system for $c^{n+1}$ that
appears in standard semi-implicit discretizations of diffusion and
is solved using a standard multigrid method. The velocity equation
in (\ref{eq:corrector}) {[}and similarly in (\ref{eq:predictor}){]}
is a much harder {}``saddle-point'' systems of linear equations
to be solved for the variables $\V v^{n+1}$ and $\pi^{n+\frac{1}{2}}$.
This time-dependent Stokes problem is solved using a Krylov iterative
solver as described in detail in Ref. \cite{NonProjection_Griffith}.
The ill-conditioning of the Stokes system is mitigated by using a
projection method (an inhomogeneous Helmholtz solve for velocity followed
by a Poisson solve for the pressure) as a preconditioner. With periodic
boundary conditions solving the Stokes system is equivalent to a projection
method, that is, to an unconstrained step for the velocities followed
by an application of the projection operator. If physical boundaries
are present, then the projection method is only an approximate solver
for the incompressible Stokes equations; however, the \textquotedbl{}splitting
error\textquotedbl{} incurred by the approximations inherent in the
projection method is corrected by the Krylov solver.

The nonlinear terms are approximated in the corrector stage using
an \emph{explicit trapezoidal rule},\begin{equation}
\M A_{\V v}^{n+\frac{1}{2}}=\frac{1}{2}\left[\M A_{\V v}(\V v^{n},c^{n})+\M A_{\V v}(\tilde{\V v}^{n+1},\tilde{c}^{n+1})\right],\label{eq:explicit_trapezoidal}\end{equation}
which is the (optimal) two-stage strong stability preserving Runge-Kutta
method \cite{SSP_Review}, and is thus generally preferable for hyperbolic
conservation laws. For the stochastic forcing terms, we employ a temporal
discretization that uses one random flux per time step,\[
\V f_{v}^{n+\frac{1}{2}}=\V f_{v}\left(\widetilde{\V W}_{\V v}^{n}\right)\mbox{ and }f_{c}^{n+\frac{1}{2}}=f_{c}\left(\tilde{c}^{n+\frac{1}{2}},\,\V W_{c}^{n}\right),\]
where $\tilde{c}^{n+\frac{1}{2}}=\left(c^{n}+\tilde{c}^{n+1}\right)/2$
but we again emphasize that the dependence of $f_{c}^{n+\frac{1}{2}}$
on the instantaneous state $\tilde{c}^{n+\frac{1}{2}}$ is not important
in the weak-noise (linearized) setting. It can be shown that this
temporal discretization is second-order weakly accurate for additive-noise
nonlinear stochastic differential equations \cite{DFDB}. More importantly,
the Crank-Nicolson method balances the numerical dissipation with
the stochastic forcing \emph{identically} in the linear setting. This
important property allows our time stepping to under-resolve the fast
dynamics of the small-wavelength fluctuations while still maintaining
the correct spectrum for the fluctuations at all scales. While this
surprising fact has already been verified (in a simplified setting)
in the Appendix of Ref. \cite{LLNS_S_k} and also in Ref. \cite{ImplicitSDEs_Weinan},
we give a different derivation in Appendix \ref{sec:CNAccuracy}.
A more detailed analysis will be presented in a forthcoming paper
\cite{DFDB}.

The \emph{linearized} equations (\ref{eq:incomp_quasi_periodic})
have additional structure that enables us to simplify the predictor-algorithm.
Firstly, the momentum equation is independent of the concentration
equation(s), $\M A_{\V v}(\V v,c)=\V 0$, and the corrector step of
the velocity equation is redundant since it simply repeats the predictor
step, $\tilde{\V v}^{n+1}=\V v^{n+1}$. Therefore, we only need to
do one Stokes solve per time step. Furthermore, only velocity enters
in the linearized concentration equation, $\M A_{c}(\V v,c)=\tilde{\M A}\V v$,
and therefore \[
\M A_{c}^{n+\frac{1}{2}}=\tilde{\M A}\tilde{\V v}^{n+\frac{1}{2}}=\tilde{\M A}\frac{\left(\V v^{n}+\V v^{n+1}\right)}{2}=\tilde{\M A}\V v^{n+\frac{1}{2}}\]
can be calculated without performing a predictor step for the concentration.
This variation of the time stepping is twice as efficient and can
be thought of a \emph{split }algorithm in which we first do a Crank-Nicolson
step for the velocity equation,\begin{equation}
\frac{\V v^{n+1}-\V v^{n}}{\D t}+\M G\pi^{n+\frac{1}{2}}=\nu\M L\left(\frac{\V v^{n+1}+\V v^{n}}{2}\right)+\rho^{-1}\V f_{v}\left(\widetilde{\V W}_{\V v}^{n}\right)\mbox{ such that }\M D\V v^{n+1}=\V 0,\label{eq:split_v}\end{equation}
and then a Crank-Nicolson step for the concentration equation using
the midpoint velocity to calculate advective fluxes,\begin{equation}
\frac{c^{n+1}-c^{n}}{\D t}=\tilde{\M A}\left(\frac{\V v^{n+1}+\V v^{n}}{2}\right)+\chi\M L\left(\frac{c^{n+1}+c^{n}}{2}\right)+\rho^{-1}f_{c}\left(c^{n},\M W_{c}^{n}\right).\label{eq:split_c}\end{equation}
Because of the special structure of the equations, the split algorithm
is equivalent to the traditional Crank-Nicolson method applied to
the \emph{coupled} velocity-concentration system in which both advection
and diffusion are treated semi-implicitly. This observation, together
with the derivation in Appendix \ref{sec:CNAccuracy}, shows that
our temporal discretization gives the correct steady-state covariances
for \emph{any time step} size $\D t$, although it does not reproduce
the correct dynamics for large $\D t$. This property will prove very
useful for the simulations of giant fluctuations reported in Section
\ref{sec:Giant-Fluctuations}.

\subsection{\label{sub:SpatialDiscretization}Spatial Discretization}

We now consider spatial discretization of the equations of fluctuating
hydrodynamics on a regular Cartesian grid, focusing on two dimensions
for notational simplicity. The spatial discretization is to be interpreted
in the finite-volume sense, that is, the value of a fluctuating field
at the center of a spatial cell of volume $\D V$ represents the average
value of the fluctuating field over the cell. We explicitly enforce
strict local conservation by using a conservative discretization of
the divergence. Specifically, the change of the average value inside
a cell can always be expressed as a sum of fluxes through each of
the faces of the cell, even if we do not explicitly write it in that
form.

Consider at first a simplified form of the stochastic advection-diffusion
equation for a scalar concentration field\begin{equation}
\partial_{t}c=\grad\cdot\left[-c\V v+\chi\grad c+\sqrt{2\chi}\,\M{\mathcal{W}}_{c}\right],\label{eq:conservative_formulation}\end{equation}
where $\V v(\V r,t)$ is a given advection velocity. We note that
for incompressible flow, we can split the stochastic stress tensor
$\M W_{v}$ into a vector $\M W_{x}$ corresponding to the flux for
$\V v_{x}$, and a vector $\M W_{y}$ corresponding to $\V v_{y}$.
We can then view the velocity equation as a constrained pair of stochastic
advection-diffusion equations of the form (\ref{eq:conservative_formulation}),
one equation for $\V v_{x}$ and another for $\V v_{y}$. We will
discuss the generalization to compressible flow in Section \ref{sub:SummarySpatial}.

The spatial discretization described in this section is to be combined
with a suitable stable temporal discretization, specifically, the
temporal discretization that we employ was described in Section \ref{sub:TemporalDiscretization}.
We consider here the limit of small time steps, $\D t\rightarrow0$,
corresponding formally to a semi-discrete {}``method of lines''
spatial discretization of the form\begin{equation}
\frac{d\V c}{dt}=\M D\left[\left(-\M U\V c+\chi\V G\V c\right)+\sqrt{2\chi/\left(\D V\,\D t\right)}\V W_{c}\right],\label{advdiff_spatial_generic}\end{equation}
where $\V c=\left\{ c_{i,j}\right\} $ is a finite-volume representation
of the random field $c(\V r,t)$. Here, $\M D$ is a conservative
discrete divergence, $\M G$ is a discrete gradient, and $\M U\equiv\M U\left(\V v\right)$
denotes a discretization of advection by the spatially-discrete velocity
field $\V v$, and $\M W_{c}$ denotes a vector of normal variates
with specifed covariance $\M C_{\M W}=\av{\V W_{c}\V W_{c}^{\star}}$.

\subsubsection{\label{sec:DFDB}Discrete Fluctuation-Dissipation Balance}

We judge the weak accuracy of the spatial discretization by comparing
the steady-state covariance of the spatially-discrete fields to the
theoretical covariance of the continuum fields in the limit $\D t\rightarrow0$
\cite{LLNS_S_k}. Ignoring for a moment constraints such as incompressibility,
at thermodynamic equilibrium the variance of the discrete fields should
be inversely proportional to $\D V$ and values in distinct cells
should be uncorrelated\begin{equation}
\M C_{\V c}=\left\langle \V c\V c^{\star}\right\rangle =S_{c,c}\,\left(\D V^{-1}\M I\right).\label{eq:C_c_model}\end{equation}
For periodic systems this means that the spectral power of each discrete
Fourier mode be equal to the continuum structure factor, $S_{c,c}=1$
for the model equation (\ref{eq:conservative_formulation}) {[}see
also (\ref{eq:S_equilibrium}){]}, independent of the wavenumber.

A spatial discretization that gives the correct equilibrium discrete
covariance is said to satisfy the \emph{discrete fluctuation-dissipation
balance} (DFDB) condition \cite{LLNS_S_k,DFDB}. The condition guarantees
that for sufficiently small time steps the statistics of the discrete
fluctuations are consistent with the continuum formulation. For larger
time steps, the difference between the discrete and continuum covariance
will depend on the order of weak accuracy of the temporal discretization
\cite{InvariantMeasureWeak}. A simple way to obtain the DFDB condition
is from the time stationarity of the covariance. For the model equation
(\ref{eq:conservative_formulation}) we obtain the linear system of
equations for the matrix $\M C_{\V c}$, \begin{equation}
\frac{d\M C_{\V c}}{dt}=\frac{d\left\langle \V c\V c^{\star}\right\rangle }{dt}=\M D\left(-\M U+\chi\V G\right)\M C_{\V c}+\M C_{\V c}\left[\M D\left(-\M U+\chi\V G\right)\right]^{\star}+2\chi\D V^{-1}\M D\M C_{\M W}\M D^{\star}=\M 0,\label{eq:DFDB_model}\end{equation}
whose solution we would like to be given by (\ref{eq:C_c_model}),
specifically, $\M C_{\V c}=\D V^{-1}\M I$. Considering first the
case of no advection, $\M U=\M 0$, we obtain the DFDB condition \begin{equation}
\M D\M G+\left(\M D\M G\right)^{\star}=-2\M D\M C_{\M W}\M D^{\star}.\label{eq:DFDB_condition}\end{equation}

Consider first the case of periodic boundary conditions. A straightforward
way to ensure the condition (\ref{eq:DFDB_condition}) is to take
the components of the random flux $\M W_{c}$ to be uncorrelated normal
variates with mean zero and unit variance, $\M C_{\M W}=\M I$, and
also choose the discrete divergence and gradient operators to be negative
adjoints of each other, $\M G=-\M D^{\star}$, just as the continuum
operators are \cite{LLNS_S_k,AMR_ReactionDiffusion_Atzberger,SELM}
(see Eq. \ref{eq:duality_cont}). Alternative approaches and the advantages
of the above {}``random flux'' approach are discussed in Ref. \cite{AMR_ReactionDiffusion_Atzberger}.
As we will demonstrate numerically in Section \ref{sec:NumericalTests},
the staggered discretization of the dissipative and stochastic terms
described below satisfies the discrete fluctuation-dissipation balance
for both compressible and incompressible flow.

In the continuum equation (\ref{eq:conservative_formulation}), the
advective term does not affect the fluctuation-dissipation balance
at equilibrium; advection simply transports fluctuations without dissipating
or amplifying them. This follows from the skew-adjoint property\begin{equation}
\int_{\Omega}w\left[\grad\cdot\left(c\V v\right)\right]d\V r=-\int_{\Omega}c\left[\grad\cdot\left(w\V v\right)\right]d\V r\mbox{ if }\grad\cdot\V v=0\mbox{ and }\V v\cdot\V n_{\partial\Omega}=0\mbox{ or }\V v\mbox{ is periodic},\label{eq:duality_cont}\end{equation}
which holds for any scalar field $w(\V r)$. In particular, choosing
$w\equiv c$ shows that for an advection equation $\partial_{t}c=-\grad\cdot\left(c\V v\right)$
the {}``energy'' $\int c^{2}\, d\V r/2$ is a conserved quantity.
To ensure that the \emph{discrete} fluctuation-dissipation balance
(\ref{eq:DFDB_model}) is satisfied, the matrix $\M D\M U\M C_{\V c}$,
or more precisely, the discrete advection operator $\M S=\M D\M U$
should be skew-adjoint, $\M S^{\star}=-\M S$. Specifically, denoting
with $\V c\cdot\V w=\sum_{i,j}c_{i,j}w_{i,j}$ the discrete dot product,
we require that for all $\V w$\begin{equation}
\V w\cdot\left[\left(\M D\M U\right)\V c\right]=-\V c\cdot\left[\left(\M D\M U\right)\V w\right]\label{eq:skew_adj_condition}\end{equation}
if the advection velocities are discretely-divergence free, $\left(\M D\V U\right)\V 1=\V 0$,
where $\V 1$ denotes a vector of all ones. Note that this last condition,
$\M S\V 1=\V 0$, ensures the desirable property that the advection
is constant-preserving, that is, advection by the random velocities
does not affect a constant concentration field.

For incompressible flow, the additional constraint on the velocity
$\M D\V v=\M 0$ needs to be taken into account when considering discrete
fluctuation-dissipation balance. In agreement with (\ref{C_vv_incompressible}),
we require that the equilibrium covariance of the discrete velocities
be\begin{equation}
\left\langle \V v\V v^{\star}\right\rangle =\rho_{0}^{-1}k_{B}T_{0}\,\left(\D V^{-1}\M{\Set P}\right),\label{eq:C_vv_nodt}\end{equation}
where $\Set P$ is the \emph{discrete projection} operator\[
\M{\Set P}=\M I-\M G\left(\M D\M G\right)^{-1}\M D=\M I-\M D^{\star}\left(\M D\M D^{\star}\right)^{-1}\M D.\]
With periodic boundary conditions, (\ref{eq:C_vv_nodt}) implies that
the discrete structure factor for velocity is $S_{\V v,\V v}=\rho_{0}^{-1}k_{B}T_{0}\,\widehat{\M{\Set P}}$.
In particular, the variance of the velocity in each cell is in agreement
with the continuum result, since $\mbox{Tr}\,\widehat{\M{\Set P}}=\mbox{Tr}\,\widehat{\M{\mathcal{P}}}=d-1$.
More generally, for non-periodic or non-uniform systems, we require
that for sufficiently small time steps all discretely-incompressible
velocity modes are equally strong at equilibrium \cite{DFDB}. In
Appendix \ref{sec:ProjectionAccuracy} we generalize the DFDB condition
(\ref{eq:DFDB_condition}) to the incompressible (constrained) velocity
equation, and show that there are no additional conditions required
from the discrete operators other than the duality condition on the
divergence and gradient operators, $\M G=-\M D^{\star}$.

\subsubsection{Staggered Grid}

A cell-centered discretization that is of the form (\ref{advdiff_spatial_generic})
and satisfies the discrete fluctuation-dissipation balance (DFDB)
condition was developed for compressible flow in Ref. \cite{LLNS_S_k}.
Extending this scheme to incompressible flow is, however, nontrivial.
In particular, imposing a strict discrete divergence-free condition
on collocated velocities has proven to be difficult and is often enforced
only approximately \cite{ApproximateProjection_I}, which is inconsistent
with (\ref{eq:C_vv_nodt}), as we explain in Appendix \ref{sec:ProjectionAccuracy}.
An alternative is to use a staggered grid or {}``MAC'' discretization,
as first employed in projection algorithms for incompressible flow
\cite{HarWel65}. In this discretization, scalars are discretized
at cell centers, i.e., placed at points $(i,j)$, while vectors (notably
velocities) are discretized on faces of the grid, placing the $x$
component at points $(i+1/2,j)$, and the $y$ component at $(i,j+1/2)$.
Such a staggered discretization is used for the fluxes in Ref. \cite{LLNS_S_k},
the main difference here being that velocities are also staggered.

In the staggered discretization, the divergence operator maps from
vectors to scalars in a locally-conservative manner,\[
\grad\cdot\V v\rightarrow\left(\M D\V v\right)_{i,j}=\D x^{-1}\left(v_{i+\frac{1}{2},j}^{(x)}-v_{i-\frac{1}{2},j}^{(x)}\right)+\D y^{-1}\left(v_{i,j+\frac{1}{2}}^{(y)}-v_{i,j-\frac{1}{2}}^{(y)}\right).\]
The discrete gradient maps from scalars to vectors, for example, for
the $x$ component:\[
\left(\grad c\right)_{x}\rightarrow\left(\M G\V c\right)_{i+\frac{1}{2},j}^{(x)}=\D x^{-1}\left(c_{i+1,j}-c_{i,j}\right).\]
It is not hard to show that with periodic boundary conditions $\M G=-\M D^{\star}$
as desired. The resulting Laplacian $\M L=\M D\M G$ is the usual
5-point Laplacian,\[
\grad^{2}c\rightarrow\left(\M L\V c\right)_{i,j}=\left[\D x^{-2}\left(c_{i-1,j}-2c_{i,j}+c_{i+1,j}\right)+\D y^{-2}\left(c_{i,j-1}-2c_{i,j}+c_{i,j+1}\right)\right],\]
which is negative definite except for the expected trivial translational
zero modes. The velocities $\V v_{x}$ and $\V v_{y}$ can be handled
analogously. For example, $\V v_{x}$ is represented on its own finite-volume
grid, shifted from the concentration (scalar) grid by one half cell
along the $x$ axis. The divergence $\M D^{(x)}$, gradient $\V G^{(x)}$
and Laplacian $\V L^{(x)}$ are the same MAC operators as for concentration,
but shifted to the $x$-velocity grid.

For the compressible equations, there is an additional dissipative
term in (\ref{eq:visc_Laplacian}) that involves $\grad\left(\grad\cdot\V v\right)$.
This term is discretized as written, $\M G\M D\V v$, which can alternatively
be expressed in conservative form. When viscosity is spatially-dependent,
the term $\grad\cdot\left(\eta\overline{\grad}\V v\right)$ should
be discretized by calculating a viscous flux on each face of the staggered
grids, interpolating viscosity as needed and using the obvious second-order
centered differences for each of the terms $\partial_{x}v_{x}$, $\partial_{x}v_{y}$,
$\partial_{y}v_{y}$ and $\partial_{y}v_{x}$. For a collocated velocity
grid the mixed derivatives $\partial_{x}v_{y}$ and $\partial_{y}v_{x}$,
and the corresponding stochastic forcing terms, do not have an obvious
face-centered discretization and require a separate treatment \cite{LLNS_S_k}.

\subsubsection{Stochastic Fluxes}

The stochastic flux $\M W_{c}$, like other vectors, is represented
on the faces of the grid, that is, $\M W_{c}$ is a vector of i.i.d.
numbers, one number for each face of the grid. To calculate the state-dependent
factor $\sqrt{c(1-c)}$ that appears in (\ref{disc_stoch_flux_covariance})
on the faces of the grid, concentration is interpolated from the cell
centers to the faces of the grid. At present, lacking any theoretical
analysis, we use a simple arithmetic average (\ref{eq:c_face_interpolation})
for this purpose.

The stochastic momentum flux $\M W_{\V v}$ is represented on the
faces of the shifted velocity grids, which for a uniform grid corresponds
to the \emph{cell centers} $(i,j)$ and the \emph{nodes} $(i+\frac{1}{2},j+\frac{1}{2})$
of the grid \cite{FluctuatingHydro_FluidOnly}. Two random numbers
need to be generated for each cell center, $W_{i,j}^{(x)}$ and $W_{i,j}^{(y)}$,
corresponding to the diagonal of the stochastic stress tensor. Two
additional random numbers need to be generated for each node of the
grid, $W_{i+\frac{1}{2},j+\frac{1}{2}}^{(x)}$ and $W_{i+\frac{1}{2},j+\frac{1}{2}}^{(y)}$,
corresponding to the off-diagonal components. In three dimensions,
the three diagonal components of the stochastic stress are represented
at the cell centers, while the six off-diagonal components are represented
at the \emph{edges} of the grid, two random numbers per edge, for
example, $W_{i+\frac{1}{2},j+\frac{1}{2},k}^{(x)}$ and $W_{i+\frac{1}{2},j+\frac{1}{2},k}^{(y)}$.

For the incompressible equations one can simply generate the different
components of $\M W_{\V v}$ as uncorrelated normal variates with
mean zero and unit variance, and obtain the correct equilibrium covariances.
Alternatively, each realization of the stochastic stress can be made
strictly symmetric and traceless as for compressible flow, as specified
in (\ref{stoch_flux_covariance}). Because of the symmetry, in practice
for each node or edge of the grid we generate only a single unit normal
variate representing the two diagonally-symmetric components. For
each cell center, we represent the diagonal components by generating
$d$ independent normal random numbers of variance $2$ and then subtracting
their average from each number. Note that for collocated velocities
a different approach is required because the diagonal and diagonally-symmetric
components of the stress tensor are not discretized on the same grid
\cite{LLNS_S_k}.

\subsubsection{Advection}

We now consider skew-adjoint discretizations of the advection operator
$\M S=\M D\M U$ on a staggered grid. This problem has been considered
in a more general context for the purpose of constructing stable methods
for turbulent flow in Ref. \cite{ConservativeDifferences_Incompressible,SkewAdjoint_lowMach};
here we focus on a simple second-order centered discretization. The
importance of the skew-adjoint condition in turbulent flow simulation
is that it leads to strict discrete energy conservation for inviscid
flow, which not only endows the schemes with long-time stability properties,
but also removes undesirable numerical dissipation. Conservation of
the discrete kinetic energy $E_{k}=\rho\left\langle \V v\cdot\V v\right\rangle /2$
is also one of the crucial ingredients for fluctuation-dissipation
balance, i.e., the requirement that the Gibbs-Boltzmann distribution
$Z^{-1}\exp\left[-E_{k}/\left(k_{B}T\right)\right]$ be the invariant
distribution of the stochastic velocity dynamics \cite{AugmentedLangevin,LB_SoftMatter_Review,SELM}.

Consider first the spatial discretization of the advective term $\M D\M U\V c$
in the concentration equation. Since divergence acts on vectors, which
are represented on the faces of the grid, $\M U\V c$ should be represented
on the faces as well, that is, $\M U$ is a linear operator that maps
from cell centers to faces, and is a consistent discretization of
the advective flux $c\V v$. If we define an advection velocity $\V u$
on the faces of the grid, and also define a concentration $\bar{\V c}$
on each face of the grid, then the advective flux can directly be
calculated on each face. For example, for the $x$ faces: \begin{equation}
\left(c\V v\right)_{x}\rightarrow\left(\M U\V c\right)_{i+\frac{1}{2},j}^{(x)}=u_{i+\frac{1}{2},j}^{(x)}c_{i+\frac{1}{2},j}.\label{eq:UC_x}\end{equation}
For concentration we can take $\V u=\V v$, since the velocity is
already represented on the faces of the scalar grid. Simple averaging
can be used to interpolate scalars from cells to faces, for example,\begin{equation}
c_{i+\frac{1}{2},j}=\frac{1}{2}\left(c_{i+1,j}+c_{i,j}\right),\label{eq:c_face_interpolation}\end{equation}
although higher-order centered interpolations can also be used \cite{LLNS_S_k}.

As discussed in Section \ref{sec:DFDB}, we require that the advection
operator be skew adjoint if $\M D\M U\V 1=\M D\V u=\V 0$. Our temporal
discretization of the incompressible equations (\ref{eq:predictor},\ref{eq:corrector})
ensures that a discretely divergence-free velocity is used for advecting
all variables. The case of compressible flow will be discussed further
in Section \ref{sub:SummarySpatial}. In the incompressible case,
$\M S=\M D\M U$ can be viewed as a second-order discretization of
the {}``skew-symmetric'' form of advection \cite{ConservativeDifferences_Incompressible}
\[
\V v\cdot\grad c+\frac{c}{2}\grad\cdot\V v=\frac{1}{2}\left[\grad\cdot\left(c\V v\right)+\V v\cdot\grad c\right].\]
Namely, using (\ref{eq:UC_x}) we obtain \[
\left(\M D\M U\V c\right)_{i,j}=\D x^{-1}\left(u_{i+\frac{1}{2},j}^{(x)}c_{i+\frac{1}{2},j}-u_{i-\frac{1}{2},j}^{(x)}c_{i-\frac{1}{2},j}\right)+\D y^{-1}\left(u_{i,j+\frac{1}{2}}^{(y)}c_{i,j+\frac{1}{2}}-u_{i,j-\frac{1}{2}}^{(y)}c_{i,j-\frac{1}{2}}\right),\]
and rewrite the $x$ term using (\ref{eq:c_face_interpolation}) as\[
\left(u_{i+\frac{1}{2},j}^{(x)}c_{i+\frac{1}{2},j}-u_{i-\frac{1}{2},j}^{(x)}c_{i-\frac{1}{2},j}\right)=\frac{1}{2}\left[\left(u_{i+\frac{1}{2},j}^{(x)}c_{i+1,j}-u_{i-\frac{1}{2},j}^{(x)}c_{i-1,j}\right)+c_{i,j}\left(u_{i+\frac{1}{2},j}^{(x)}-u_{i-\frac{1}{2},j}^{(x)}\right)\right],\]
and similarly for the $y$ term, to obtain \begin{equation}
\left(\M D\M U\V c\right)_{i,j}=\left(\M S\V c\right)_{i,j}=\left(\widetilde{\M S}\V c\right)_{i,j}+\frac{1}{2}c_{i,j}\left(\M D\V u\right)_{i,j},\label{eq:adv_S_tilde}\end{equation}
where $\widetilde{\M S}$ is a centered discretization of $\left[\grad\cdot\left(c\V v\right)+\V v\cdot\grad c\right]/2$,\begin{equation}
\left(\widetilde{\M S}\V c\right)_{i,j}=\frac{1}{2}\left[\D x^{-1}\left(u_{i+\frac{1}{2},j}^{(x)}c_{i+1,j}-u_{i-\frac{1}{2},j}^{(x)}c_{i-1,j}\right)+\D y^{-1}\left(u_{i,j+\frac{1}{2}}^{(y)}c_{i,j+1}-u_{i,j-\frac{1}{2}}^{(y)}c_{i,j-1}\right)\right].\label{eq:S_tilde}\end{equation}
Since the advection velocity is discretely divergence free, $\M S=\widetilde{\M S}$.

It is not hard to show that $\widetilde{\M S}$ is skew-adjoint. Consider
the $x$ term in $\left[\widetilde{\M S}\V c\right]\cdot\V w$, and,
assuming periodic boundary conditions, shift the indexing from $i$
to $i-1$ in the first sum and from $i$ to $i+1$ in the second sum,
to obtain\[
\sum_{i,j}w_{i,j}\left(u_{i+\frac{1}{2},j}^{(x)}c_{i+1,j}-u_{i-\frac{1}{2},j}^{(x)}c_{i-1,j}\right)=-\sum_{i,j}c_{i,j}\left(u_{i+\frac{1}{2},j}^{(x)}w_{i+1,j}-u_{i-\frac{1}{2},j}^{(x)}w_{i-1,j}\right).\]
Therefore, $\widetilde{\M S}$ is skew-adjoint, $\left(\widetilde{\M S}\V c\right)\cdot\V w=-\V c\cdot\left(\widetilde{\M S}\V w\right)$.
A similar transformation can be performed with slip or stick boundary
conditions as well. These calculations show that (\ref{eq:skew_adj_condition})
holds and thus the discrete advection operator is skew-adjoint, as
desired. Note that the additional terms in (\ref{eq:LLNS_incomp_c})
due to the Soret effect can be included by advecting concentration
with the effective velocity $\V v_{\text{adv}}=\V v-\chi S_{T}\grad T$.

The same approach we outlined above for concentration can be used
to advect the velocities as well. Each velocity component lives on
its own staggered grid and advection velocities are needed on the
faces of the shifted grid, which in two dimensions corresponds to
the cell centers and the nodes of the grid. The velocity $\V v_{x}$
is advected using an advection velocity field $\V u^{(x)}$ that is
obtained via a second-order interpolation of $\V v$,\begin{eqnarray*}
\left(u_{x}^{(x)}\right)_{i,j} & = & \frac{1}{2}\left(v_{i-\frac{1}{2},j}^{(x)}+v_{i+\frac{1}{2},j}^{(x)}\right)\\
\left(u_{y}^{(x)}\right)_{i+\frac{1}{2},j+\frac{1}{2}} & = & \frac{1}{2}\left(v_{i,j+\frac{1}{2}}^{(y)}+v_{i+1,j+\frac{1}{2}}^{(y)}\right),\end{eqnarray*}
and similarly for the other components. It is not hard to verify that
the advection velocity $\V u^{(x)}$ is discretely divergence-free
if $\V v$ is:\[
\left(\M D^{(x)}\V u^{(x)}\right)_{i+\frac{1}{2},j}=\frac{1}{2}\left[\left(\M D\V v\right)_{i,j}+\left(\M D\V v\right)_{i+1,j}\right],\]
showing that $\M D^{(x)}\V u^{(x)}=\V 0$ if $\M D\V v=\V 0$. Therefore,
the shifted advection operator $\M S^{(x)}=\M D^{(x)}\V U^{(x)}$
is also skew-adjoint, as desired.

\subsubsection{\label{sub:SummarySpatial}Compressible Equations}

It is instructive at this point to summarize our spatial discretization
of the incompressible equations (\ref{eq:LLNS_incomp_v},\ref{eq:LLNS_incomp_c}),
before turning to the compressible equations. The concentration equation
(\ref{eq:LLNS_incomp_c}) is discretized as\begin{equation}
\frac{d\V c}{dt}=-\M D\M U\V c+\chi\M D\V G\V c+\M D\M{\Psi},\label{eq:c_spatial}\end{equation}
where $\M U$ is given by (\ref{eq:UC_x}) with advection velocity
$\V u=\V v-\chi S_{T}\grad T$. For the $x$ component of the velocity
we use the spatial discretization \[
\frac{d\V v_{x}}{dt}+\left(\M G\V{\pi}\right)_{x}=-\M D^{(x)}\V U^{(x)}\V v_{x}+\eta\M D^{(x)}\M G^{(x)}\V v_{x}+\rho^{-1}\M D^{(x)}\M{\Sigma}^{(x)},\]
and similarly for the other components, and the pressure ensures that
$\M D\V v=\V 0$.

Our staggered spatial discretization of the compressible equations
(\ref{eq:LLNS_rho_simp},\ref{eq:LLNS_comp_v_simp},\ref{eq:LLNS_comp_c_simp})
is closely based on the discretization described above for the incompressible
equations. An important difference is that for compressible flow we
use the conservative form of the equations, that is, we use the mass
density $\rho$, the momentum density $\V j=\rho\V v$ and the partial
mass density $\rho_{1}=c\rho$ as variables. The momentum densities
are staggered with respect to the mass densities. Staggered velocities
are defined by interpolating density from the cell centers to the
faces of the grid, for example,\[
\V v_{i+\frac{1}{2},j}^{(x)}=\V j_{i+\frac{1}{2},j}^{(x)}/\rho_{i+\frac{1}{2},j}=2\V j_{i+\frac{1}{2},j}^{(x)}/\left(\rho_{i+1,j}+\rho_{i,j}\right),\]
which implies that $\M D\V j=\M D\V U\V{\rho}$.

The density equation (\ref{eq:LLNS_rho_simp}) is discretized spatially
as\begin{equation}
\frac{d\V{\rho}}{dt}=-\M D\V U\V{\rho},\label{eq:rho_spatial}\end{equation}
while for the concentration equation (\ref{eq:LLNS_comp_c_simp})
we use\begin{equation}
\frac{d\V{\rho}_{1}}{dt}=-\M D\M U\V{\rho_{1}}+\rho_{0}\chi_{0}\M D\V G\V c+\M D\M{\Psi},\label{eq:rho1_spatial}\end{equation}
where we assume that $\rho\chi=\rho_{0}\chi_{0}$ is constant. For
the $x$ component of the momentum density we use\begin{equation}
\frac{d\V j_{x}}{dt}=-\M D^{(x)}\V U^{(x)}\V j_{x}-c_{T}^{2}\left(\M G\V{\rho}\right)_{x}+\eta\M D^{(x)}\M G^{(x)}\V v_{x}+\left(\zeta+\frac{\eta}{3}\right)\left(\M G\M D\V v\right)_{x}+\M D^{(x)}\M{\Sigma}^{(x)},\label{eq:jx_spatial}\end{equation}
and similarly for the other components. The spatio-temporal discretization
ensures strict local conservation of $\rho$, $\V j$ and $\rho_{1}$.

The discretization (\ref{eq:rho_spatial},\ref{eq:rho1_spatial},\ref{eq:jx_spatial})
satisfies discrete-fluctuation dissipation balance at equilibrium,
specifically, the equilibrium covariances of velocity and density
are $\av{\V v\V v^{\star}}=\rho_{0}^{-1}k_{B}T_{0}\,\M I$ and $\av{\V{\rho}\V{\rho}^{\star}}=\rho_{0}k_{B}T_{0}/c_{T}^{2}\,\M I$,
in agreement with the continuum spectra given in (\ref{eq:S_equilibrium}).
Linearizing the semi-discrete density equation (\ref{eq:rho_spatial})
around an equilibrium state $\left(\V{\rho},\V v\right)=\left(\V{\rho}_{0}+\d{\V{\rho}},\,\V v_{0}+\d{\V v}\right)$
with $\M D\V v_{0}=\M 0$ gives\[
\frac{d\left(\d{\V{\rho}}\right)}{dt}+\widetilde{\M S}_{0}\left(\d{\V{\rho}}\right)=-\rho_{0}\left[\M D\left(\delta\V v\right)\right].\]
Recall that the operator $\widetilde{\M S}_{0}$, defined by (\ref{eq:S_tilde})
with $\V u=\V v_{0}$, is skew-adjoint, and the fluctuations in density
are thus controlled by the coupling with the velocity fluctuations.
For simplicity, consider this coupling for the case of a fluid at
rest, $\V v_{0}=\V 0$ and thus $\d{\V j}=\rho_{0}\left(\d{\V v}\right)$.
Linearizing the momentum update (\ref{eq:jx_spatial}) and focusing
on the coupling with the density fluctuations, we obtain\[
\frac{d\left(\d{\V v}\right)}{dt}+\mbox{advection}=-\rho_{0}^{-1}c_{T}^{2}\left[\M G\left(\d{\V{\rho}}\right)\right]+\mbox{dissipation and forcing}.\]
Fluctuation-dissipation balance requires the skew-symmetry property
$\M L_{\rho,\V v}\av{\V v\V v^{\star}}=-\av{\V{\rho}\V{\rho}^{\star}}\M L_{\V v,\rho}^{\star}$,
where $\M L_{\rho,\V v}=-\rho_{0}\M D$ the operator in front of $\delta\V v$
in the density equation, and $\M L_{\V v,\rho}=-c_{T}^{2}\M G$ is
the operator in front of $\d{\V{\rho}}$ in the velocity equation.
This skew-symmetry requirement is satisfied because of the key duality
property $\M D=-\M G^{\star}$. This demonstrates the importance of
the duality between the discrete divergence and gradient operators,
not just for a single advection-diffusion equation, but also for coupling
between the different fluid variables. In future work, we will explore
generalizations of the concept of skew-adjoint discrete advection
to the nonlinear compressible equations \cite{OttingerBook,SkewAdjoint_lowMach}.

\subsubsection{\label{sub:BoundaryConditions}Boundary Conditions}

Non-periodic boundary conditions, specifically, Neumann or Dirichlet
physical boundaries, can be incorporated into the spatial discretization
by modifying the discrete divergence, gradient and Laplacian operators
near a boundary. This needs to be done in a way that not only produces
an accurate and robust deterministic scheme, but also ensures fluctuation-dissipation
balance even in the presence of boundaries. Here we extend the approach
first suggested in an Appendix in Ref. \cite{DSMC_Hybrid} to the
staggered grid. It can be shown that the inclusion of the (discrete)
incompressibility constraint does not affect the fluctuation-dissipation
balance when an unsplit Stokes solver is employed in the temporal
integrator \cite{DFDB}.

We assume that the physical boundary is comprised of faces of the
grid. Since only the direction perpendicular to the wall is affected,
we focus on a one-dimensional system in which there is a physical
boundary between cells $1$ and $0$. For the component of velocity
perpendicular to the wall, some of the grid points are on the physical
boundary itself and those values are held fixed and not included as
independent degrees of freedom. For the second-order spatial discretization
that we employ no values in cells outside of the physical domain are
required. Therefore, no special handling at the boundary is needed.

For cell-centered quantities, such as concentration and components
of the velocity parallel to the wall, the boundary is half a cell
away from the cell center, that is, the boundary is staggered. In
this case we use the same discrete operators near the boundaries as
in the interior of the domain, using\emph{ ghost cells} extending
beyond the boundaries to implement the finite-difference stencils
near the boundaries. One can think of this as a modification of the
stencil of the Laplacian operator $\M L$ near boundaries, specifically,
when boundaries are present the dissipative operator $\M L\neq\M D\M G$
but rather $\M L=\M D\tilde{\M G}$, where $\tilde{\M G}$ is a modified
gradient. Repeating the calculation in (\ref{eq:DFDB_model}) for
the spatially-discretized model equation\[
\frac{d\V c}{dt}=\chi\M L\V c+\sqrt{2\chi/\left(\D V\,\D t\right)}\M D\V W.\]
leads to a generalization of the DFDB condition (\ref{eq:DFDB_condition}),
assuming $\M L^{\star}=\M L$, \begin{equation}
\chi\M L\M C_{\V c}+\chi\M C_{\V c}\M L^{\star}=2\chi\D V^{-1}\M L=-2\chi\D V^{-1}\M D\M C_{\M W}\M D^{\star}\mbox{\quad\ensuremath{\Rightarrow}\quad}\M L=-\M D\M C_{\M W}\M D^{\star}.\label{eq:DFDB_wall}\end{equation}

Consider first a Neumann condition on concentration, $\partial c(0)/\partial x=0$.
This means that a no-flux condition is imposed on the boundary, and
therefore for consistency with physical conservation the stochastic
flux on the boundary should also be set to zero, $W_{\frac{1}{2}}=0$.
The ghost cell value is set equal to the value in the neighboring
interior cell (reflection), $c_{0}=c_{1}$, leading to\begin{equation}
\left(\M D\V W\right)_{1}=\D x^{-1}\, W_{\frac{3}{2}},\quad\left(\tilde{\M G}\V c\right)_{\frac{1}{2}}=0,\quad\left(\M L\V c\right)_{1}=\D x^{-2}\left(c_{2}-c_{1}\right).\label{eq:Neumann_wall}\end{equation}
If we exclude points on the boundary from the domain of the divergence
operator, which is also the range (image) of the gradient operator,
then it is not hard to see that the duality condition $\M D^{\star}=-\tilde{\M G}$
continues to hold. We can therefore continue to use uncorrelated unit
normal variates for the stochastic fluxes not on the boundary, $\M C_{\M W}=\M I$
in (\ref{eq:DFDB_wall}).

If a Dirichlet condition $c(0)=0$ is imposed, then the ghost cell
value is obtained by a linear extrapolation of the value in the neighboring
interior cell (inverse reflection), $c_{0}=-c_{1}$, leading to\begin{equation}
\left(\M D\V W\right)_{1}=\D x^{-1}\left(W_{\frac{3}{2}}-W_{\frac{1}{2}}\right),\quad\left(\tilde{\M G}\V c\right)_{\frac{1}{2}}=\D x^{-1}\left(2c_{1}\right),\quad\left(\M L\V c\right)_{1}=\D x^{-2}\left(c_{2}-3c_{1}\right).\label{eq:Dirichlet_wall}\end{equation}
The duality condition is no longer satisfied, $\M D^{\star}\neq-\tilde{\M G}$,
but it is not hard to show that the fluctuation-dissipation balance
condition (\ref{eq:DFDB_wall}) can be satisfied by simply doubling
the variance of the stochastic flux on the boundary, $\av{W_{\frac{1}{2}}W_{\frac{1}{2}}^{\star}}=2$.
Note that the Laplacian (\ref{eq:Dirichlet_wall}) is not formally
second-order accurate at the boundary, however, its normal modes (eigenvectors)
can be shown to correspond exactly to the normal modes of the continuum
Laplacian and have decay rates (eigenmodes) that are second-order
accurate in $\D x^{2}$, and in practice pointwise second-order accuracy
is observed even next to the boundary. Formal second-order local accuracy
can be obtained by using a quadratic extrapolation for the ghost cell,
$c_{0}=-2u_{1}+u_{2}/3$ and $\left(\M L\V c\right)_{1}=\D x^{-2}\left(4c_{2}/3-4c_{1}\right)$,
however, this requires a more complicated handling of the stochastic
fluxes near the boundary as well.

In summary, the only change required to accommodate physical boundaries
is to set the variance of stochastic fluxes on a physical boundary
to zero (at Neumann boundaries), or to twice that used for the interior
faces (at Dirichlet boundaries). For density in compressible flows,
the ghost cell values are generated so that the pressure in the ghost
cells is equal to the pressure in the neighboring interior cell, which
ensures that there is no unphysical pressure gradient in the momentum
equation across the interface. There is also no stochastic mass flux
through faces on the boundary independent of the type of boundary
condition at the wall. For incompressible flow the gradient of pressure
is discretized as $\M G\pi=-\M D^{\star}\pi$ even in the presence
of stick or slip boundary conditions for velocity; more complicated
velocity-stress or open \cite{Delgado:08} boundary conditions are
simple to handle (in principle) with the projection-preconditioner
solvers.

\section{\label{sec:NumericalTests}Implementation and Numerical Tests}

We now describe in more detail our implementations of the spatio-temporal
discretizations described in Section \ref{sec:SpatioTemporal}, and
provide numerical evidence of their ability to reproduce the correct
fluctuation spectrum in uniform flows with periodic boundary conditions.
A less trivial application with non-periodic boundaries is studied
in Section \ref{sec:Giant-Fluctuations}.

We consider here a uniform periodic system in which there is a steady
background (mean) flow of velocity $\V v_{0}$. Unlike the continuum
formulation, the discrete formulation is not Galilean-invariant under
such uniform motion and the covariance of the discrete fluctuations
is affected by the magnitude of $\V v_{0}$. The stability and accuracy
of the spatio-temporal discretization is controlled by the dimensionless
CFL numbers\[
\alpha=\frac{V\D t}{\D x},\mbox{\quad}\beta=\frac{\nu\D t}{\D x^{2}},\mbox{ and }\beta_{c}=\frac{\chi\D t}{\D x^{2}},\]
where $V=c_{T}$ (isothermal speed of sound) for low Mach number compressible
flow, and $V=\norm{\V v_{0}}_{\infty}$ for incompressible flow, and
typically $\chi\ll\nu$. The explicit handling of the advective terms
places a stability condition $\alpha\lesssim1$, and the explicit
handling of diffusion in the compressible flow case requires $\beta,\beta_{c}\leq1/2^{d}$,
where $d$ is the dimensionality. The strength of advection relative
to dissipation is measured by the cell Reynolds number $r=\alpha/\beta=V/\left(\nu\D x\right)$.

To characterize the weak accuracy of our methods we examine the discrete
Fourier spectra of the fluctuating fields at equilibrium, and compare
them to the continuum theory discussed in Section \ref{sub:ContinuumSpectra}
for all discrete wavenumbers $\V k$. We use subscripts to denote
which pair of variables is considered, and normalize each covariance
so that for self-correlations we report the relative error in the
variance, and for cross-correlations we report the correlation coefficient
between the two variables. For example, the non-dimensionalized static
structure factor for concentration is\[
\tilde{S}_{c,c}=\frac{\left\langle \hat{c}\hat{c}^{\star}\right\rangle }{\D V^{-1}S_{c,c}}=\frac{\D V}{M\rho_{0}^{-1}\, c_{0}(1-c_{0})}\left\langle \hat{c}\hat{c}^{\star}\right\rangle ,\]
where $\hat{c}(\V k)$ is the discrete Fourier transform of the concentration.
Note that an additional factor equal to the total number of cells
may be needed in the numerator depending on the exact definition used
for the discrete Fourier transform \cite{LLNS_S_k}. Similarly, the
cross-correlations between different variables need to be examined
as well, such as for example,\[
\tilde{\V S}_{c,\V v}=\frac{\D V}{\sqrt{\left[M\rho_{0}^{-1}\, c_{0}(1-c_{0})\right]\left(\rho_{0}^{-1}k_{B}T_{0}\right)}}\left\langle \hat{c}\hat{\V v}^{\star}\right\rangle .\]
For staggered variables the shift between the corresponding grids
should be taken into account as a phase shift in Fourier space, for
example, $\exp\left(k_{x}\D x/2\right)$ for $v_{x}$. For a perfect
scheme, $\tilde{S}_{c,c}=1$ and $\tilde{\V S}_{c,\V v}=\V 0$ for
all wavenumbers, and discrete fluctuation-dissipation balance in our
discretization ensures this in the limit $\D t\rightarrow0$. Our
goal will be to quantify the deviations from {}``perfect'' for several
methods, as a function of the dimensionless numbers $\alpha$ and
$\beta$.

\subsection{\label{sub:Incompressible-Solver}Incompressible Solver}

We have implemented the incompressible scheme described in Sections
\ref{sub:IncompressibleTemporal} and \ref{sub:SpatialDiscretization}
using the IBAMR software framework \cite{IBAMR}, an open-source library
for developing fluid-structure interaction models that use the immersed
boundary method. The IBAMR framework uses SAMRAI \cite{HornungSAMRAI06}
to manage Cartesian grids in parallel, and it uses PETSc \cite{PETSc}
to provide iterative Krylov solvers. The majority of the computational
effort in the incompressible solver is spent in the linear solver
for the Stokes system; in particular, in the projection-based preconditioner,
the application of which requires solving a linear Poisson system
for the pressure, and a modified linear Helmoltz system for the velocities
and the concentrations \cite{NonProjection_Griffith}. For small viscous
CFL numbers $\beta\ll1$ the Poisson solver dominates the cost, however,
for $\beta\gg1$ the two linear systems become similarly ill-conditioned
and require a good preconditioner themselves. We employ the \emph{hypre}
library \cite{hypre} to solve the linear systems efficiently using
geometric multigrid solvers.

For incompressible flow, one could directly compare the spectrum of
the velocities $\left\langle \hat{\V v}\hat{\V v}^{\star}\right\rangle $
to the spectrum of the discrete projection operator $\M{\Set P}$
(see Section \ref{sec:DFDB}). It is, however, simpler and more general
to instead examine the equilibrium covariance of the discrete modes
forming an orthonormal basis for the space of discretely divergence
free modes. The amplitude of each mode should be unity for all wavenumbers,
even if there are physical boundaries present, making it easy to judge
the accuracy at different wavenumbers. For periodic boundary conditions
a discretely-orthogonal basis is obtained by replacing the wavenumber
$\V k=\left(k_{x},k_{y},k_{z}\right)$ in (\ref{eq:incomp_basis_1},\ref{eq:incomp_basis_2},\ref{eq:incomp_basis_3})
by the effective wavenumber $\tilde{\V k}$ that takes into account
the centered discretization of the projection operator, for example,\begin{equation}
\tilde{k}_{x}=\frac{\exp\left(ik_{x}\D x/2\right)-\exp\left(-ik_{x}\D x/2\right)}{i\D x}=k_{x}\frac{\sin\left(k_{x}\D x/2\right)}{\left(k_{x}\D x/2\right)}.\label{eq:k_x_tilde}\end{equation}
Our temporal discretization ensures that the discrete velocities are
discretely divergence free, that is, $\av{\hat{v}_{1}\hat{v}_{1}^{\star}}=0$
to within the tolerance of the linear solvers used for the Stokes
system. For a perfect scheme, the dimensionless structure factor\[
\tilde{S}_{\V v}^{(2)}=\frac{\D V}{\rho_{0}^{-1}k_{B}T_{0}}\av{\hat{v}_{2}\hat{v}_{2}^{\star}},\]
and analogously $\tilde{S}_{\V v}^{(3)}$ (in three dimensions) would
be unity for all wavenumbers, while $\tilde{S}_{\V v}^{(2,3)}\sim\av{\hat{v}_{2}\hat{v}_{3}^{\star}}$
would be zero.

\begin{figure*}
\begin{centering}
\includegraphics[width=0.45\textwidth]{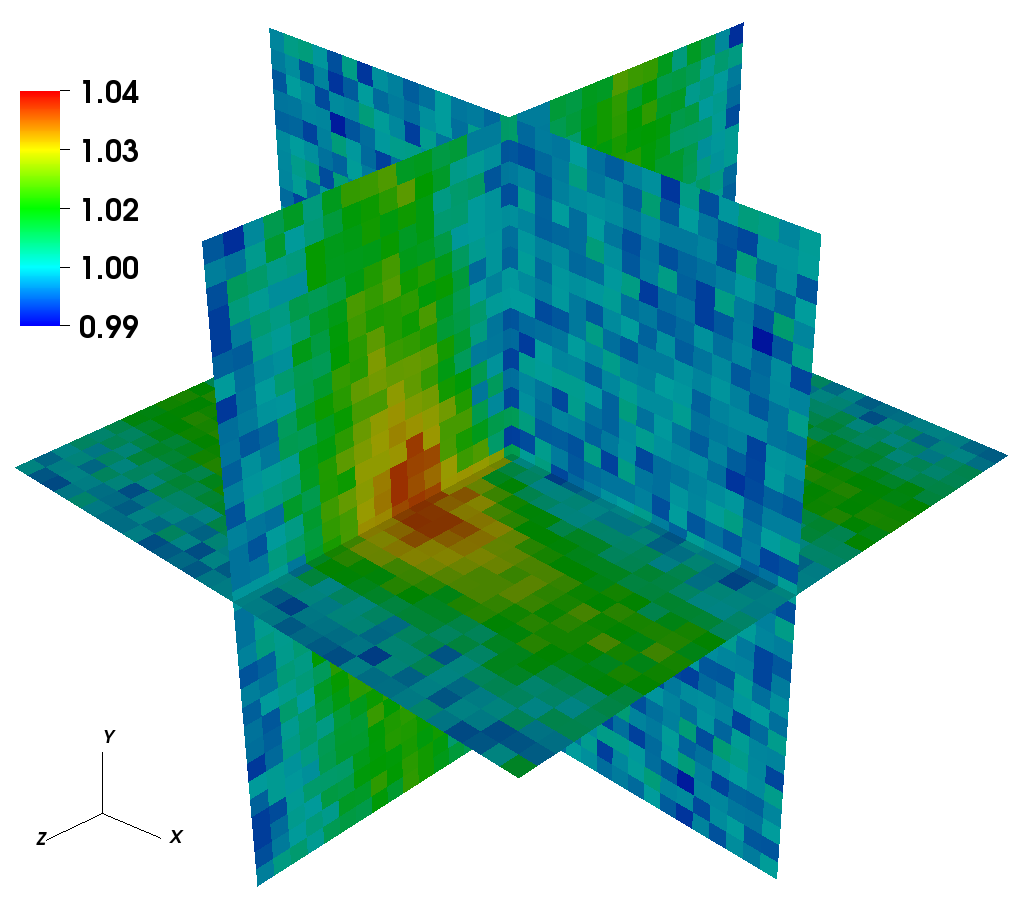}\includegraphics[width=0.45\textwidth]{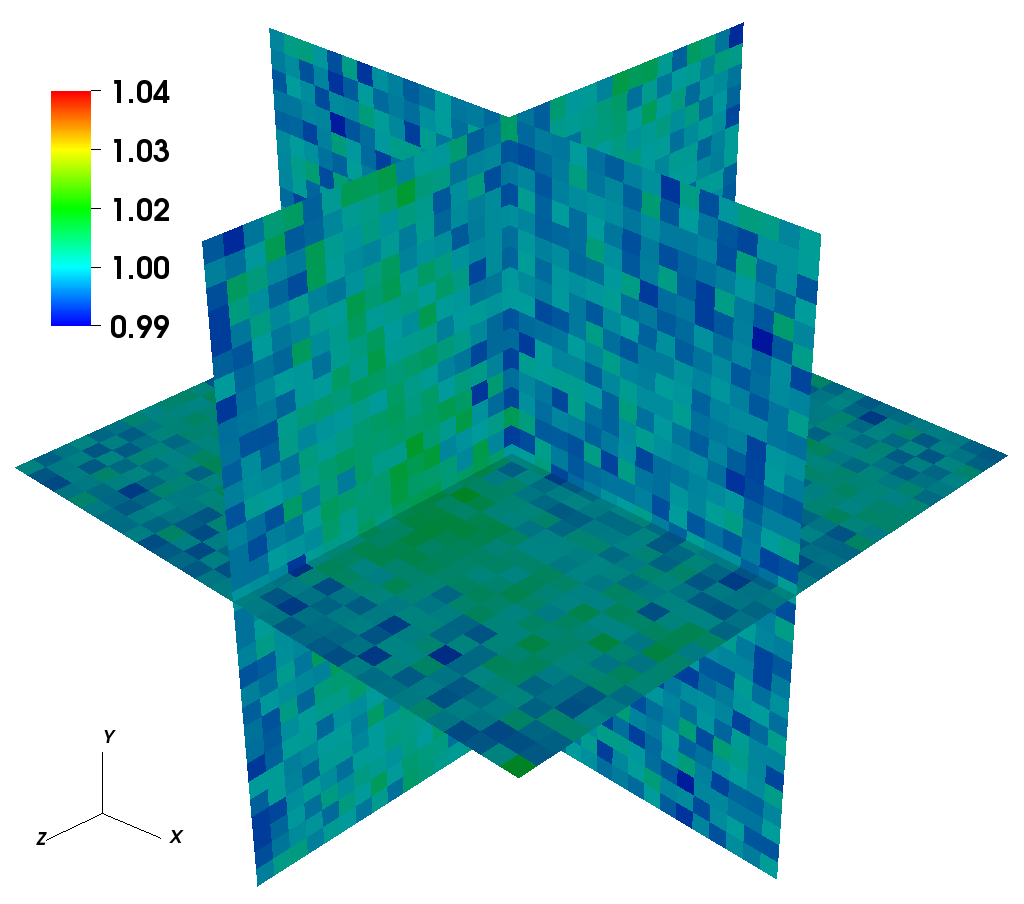}
\par\end{centering}

\caption{\label{fig:S_vv_incomp}Spectral power of the first solenoidal mode
for an incompressible fluid, $\tilde{S}_{\V v}^{(2)}\left(k_{x},k_{y},k_{z}\right)$,
as a function of the wavenumber (ranging from $0$ to $\pi/\D x$
along each axes), for a periodic system with $32^{3}$ cells. A uniform
background flow along the $z$ axis is imposed. The left panel is
for a time step $\alpha=0.5$, and the right for $\alpha=0.25$. Though
not shown, we find that $\tilde{S}_{\V v}^{(3)}$ and $\tilde{S}_{c,c}$
are essentially identical, and both the real and imaginary parts of
the cross-correlation $\tilde{S}_{\V v}^{(2,3)}$ vanish to within
statistical accuracy.}
\end{figure*}

Note that for a system at equilibrium, $\grad\bar{c}=\V 0$, the linearized
velocity equation and the concentration equation (\ref{eq:incomp_quasi_periodic})
are uncoupled and thus $\tilde{\V S}_{c,\V v}=\V 0$. Observe that
the same temporal discretization is used for the velocity equation,
projected onto the space of discretely divergence-free vector fields
consistent with the boundary conditions, and for the concentration
equation. Therefore, it is sufficient to present here numerical results
for only one of the self-correlations $\tilde{S}_{\V v}^{(2)}$, $\tilde{S}_{\V v}^{(3)}$
and $\tilde{S}_{c,c}$. In Fig. \ref{fig:S_vv_incomp} we show $\tilde{S}_{\V v}^{(2)}$
as a function of the wavenumber $\V k$ in three dimensions for a
cell Reynolds number $r=1$ and an advective CFL number $\alpha=0.5$
and $\alpha=0.25$. Even for the relatively large time step, the deviation
from unity is less than 5\%, and as $\alpha\rightarrow0$ it can be
shown theoretically and observed numerically that the correct covariance
is obtained at all wavenumbers.

\begin{figure*}
\centering{}\includegraphics[width=0.5\textwidth]{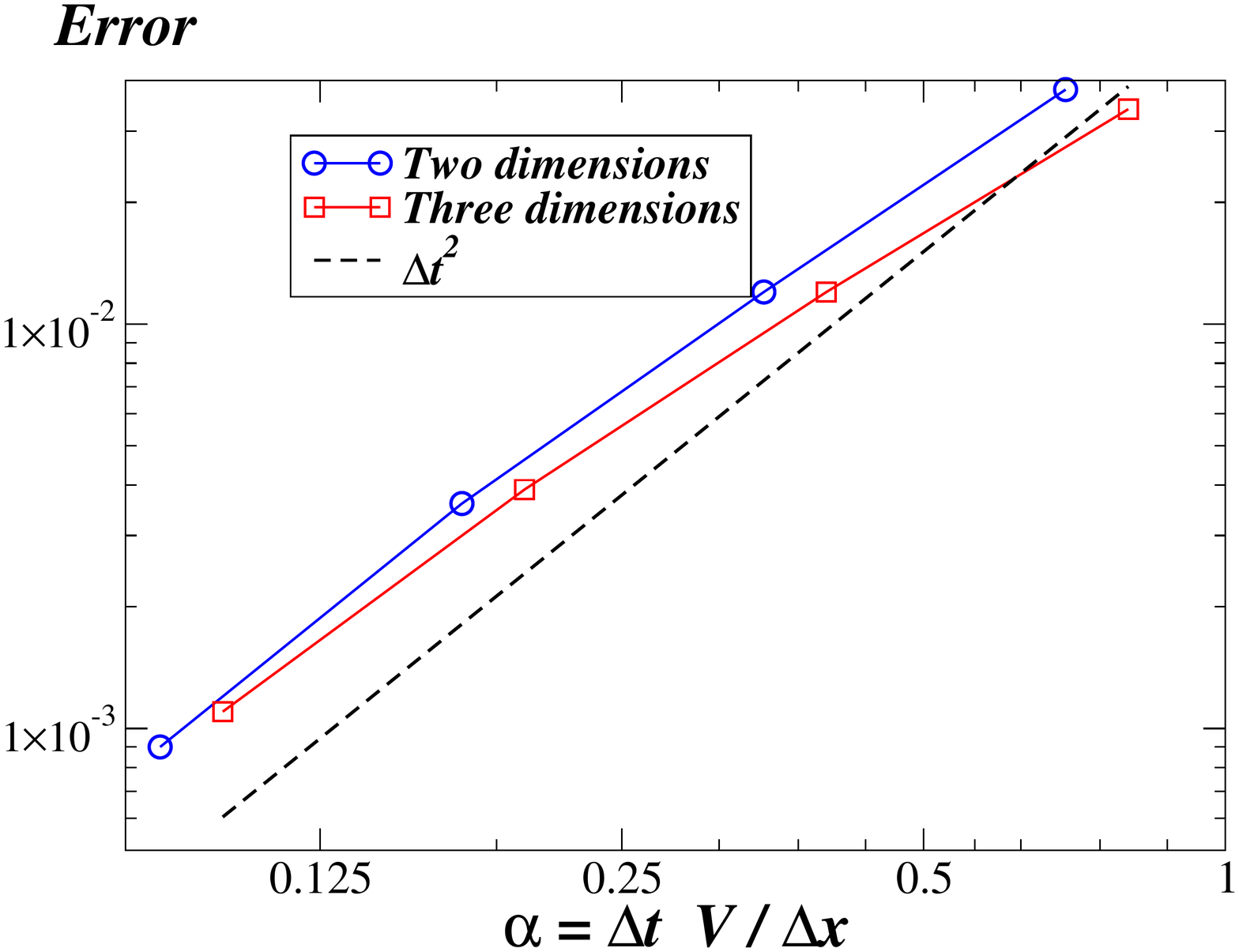}\includegraphics[width=0.5\textwidth]{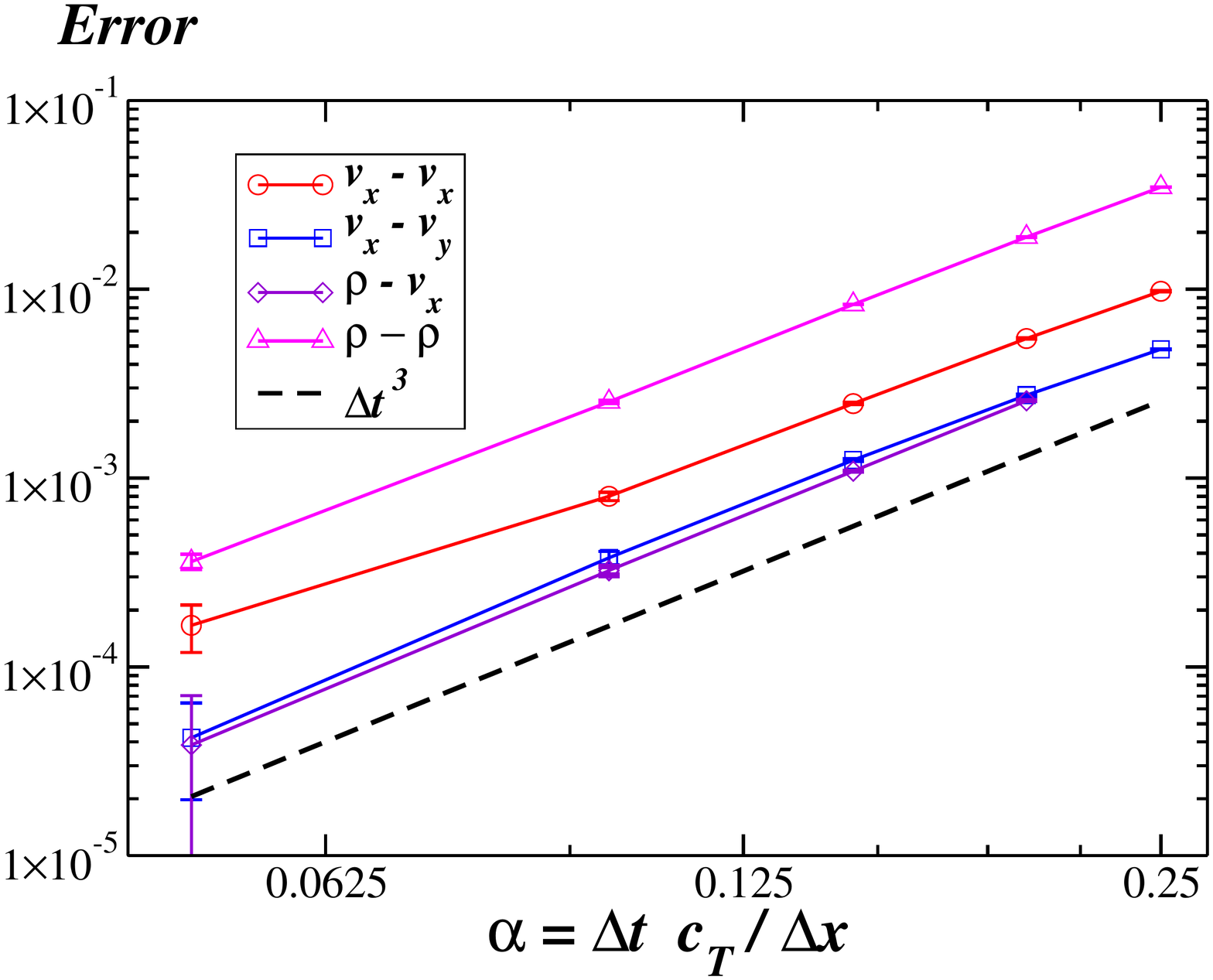}
\caption{\label{fig:error_vs_alpha}(Left) Relative error in the equilibrium
variance of velocity (or, equivalently, concentration) for several
time steps, as obtained using our incompressible code with a background
flow velocity $\V v_{0}=\left(\sqrt{3},2\right)/2$ corresponding
to cell Reynolds number $r=\sqrt{3}/2$ in two dimensions, and $\V v_{0}=\left(1,1/3,1/3\right)$
corresponding to $r=1$ in three dimensions, for a grid of size $32^{2}$
and $32^{3}$ cells, respectively. The theoretical order of convergence
$O(\D t^{2})$ is shown for comparison. Error bars are on the order
of the symbol size. (Right) Normalized covariance of the discrete
velocities and densities compared to the theoretical expectations,
using the parameters reported in the caption of Fig. \ref{fig:S_rho_v_comp}.
The value reported is the relative error of the variance of a variable
or the correlation coefficient between pairs of variables, see legend.
The theoretical order of convergence $O(\D t^{3})$ is shown for comparison.
Error bars are indicated but are smaller than the symbol size except
for the smallest time step.}
\end{figure*}

Theoretical analysis suggests that the error in the discrete covariance
vanishes with time step and the background velocity as $O(\alpha^{2})\sim O\left(V^{2}\D t^{2}\right)$
for both velocity and concentration \cite{DFDB}. In the left panel
of Fig. \ref{fig:error_vs_alpha} we show the observed relative error
in the variance of the discrete velocity as a function of $\alpha$,
confirming the predicted quadratic convergence. As expected, identical
results are obtained for concentration as well. These numerical results
confirm that our spatial discretization satisfies discrete fluctuation-dissipation
balance and the temporal discretization is weakly second-order accurate.

\subsection{\label{sub:Compressible-Solver}Compressible Solver}

Unlike the incompressible method, which requires complex linear solvers
and preconditioners, the explicit compressible scheme is very simple
and easy to parallelize on Graphics Processing Units (GPUs). Our implementation
is written in the CUDA programming environment, and is three-dimensional
with the special case of $N_{z}=1$ cell along the $z$ axes corresponding
to a quasi two-dimensional system. In our implementation we create
one thread per cell, and each thread only writes to the memory address
associated with its cell and only accesses the memory associated with
its own and neighboring cells. This avoids concurrent writes and costly
synchronizations between threads, facilitating efficient execution
on the GPU. Further efficiency is gained by using the GPU texture
unit to perform some of the simple computations such as evaluating
the equation of state. Our GPU code running in a NVIDIA GeForce GTX
480 is about 4 times faster (using double precision) than a compressible
CPU-based code \cite{LLNS_S_k} running on 32 AMD cores using MPI.
Note that with periodic boundary conditions the velocity and the pressure
linear systems in the incompressible formulation decouple and Fast
Fourier Transforms could be used to solve them efficiently. We have
used this to also implement the incompressible algorithm on a GPU
by using the NVIDIA FFT library as a Poisson/Helmholtz solver. We
emphasize, however, that this approach is applicable only to the case
of periodic boundary conditions.

\begin{figure*}
\begin{centering}
\includegraphics[width=0.45\textwidth]{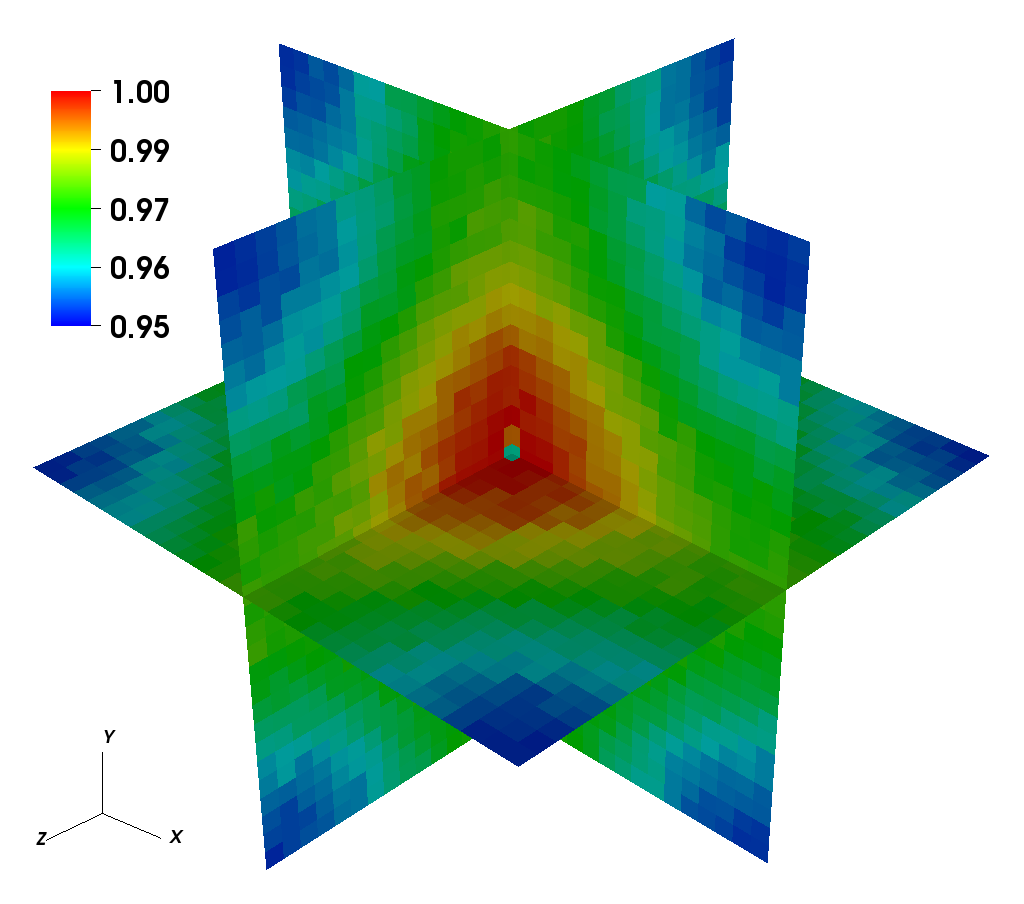}\includegraphics[width=0.45\textwidth]{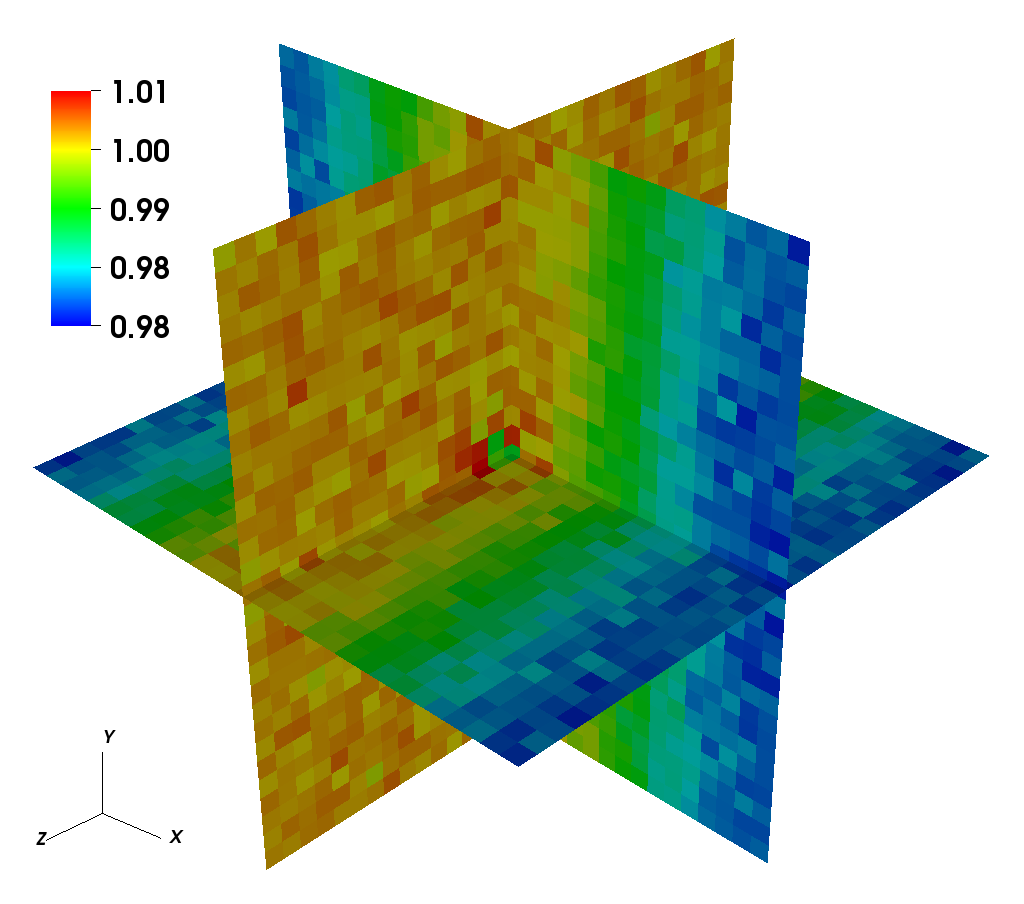}
\par\end{centering}

\begin{centering}
\includegraphics[width=0.45\textwidth]{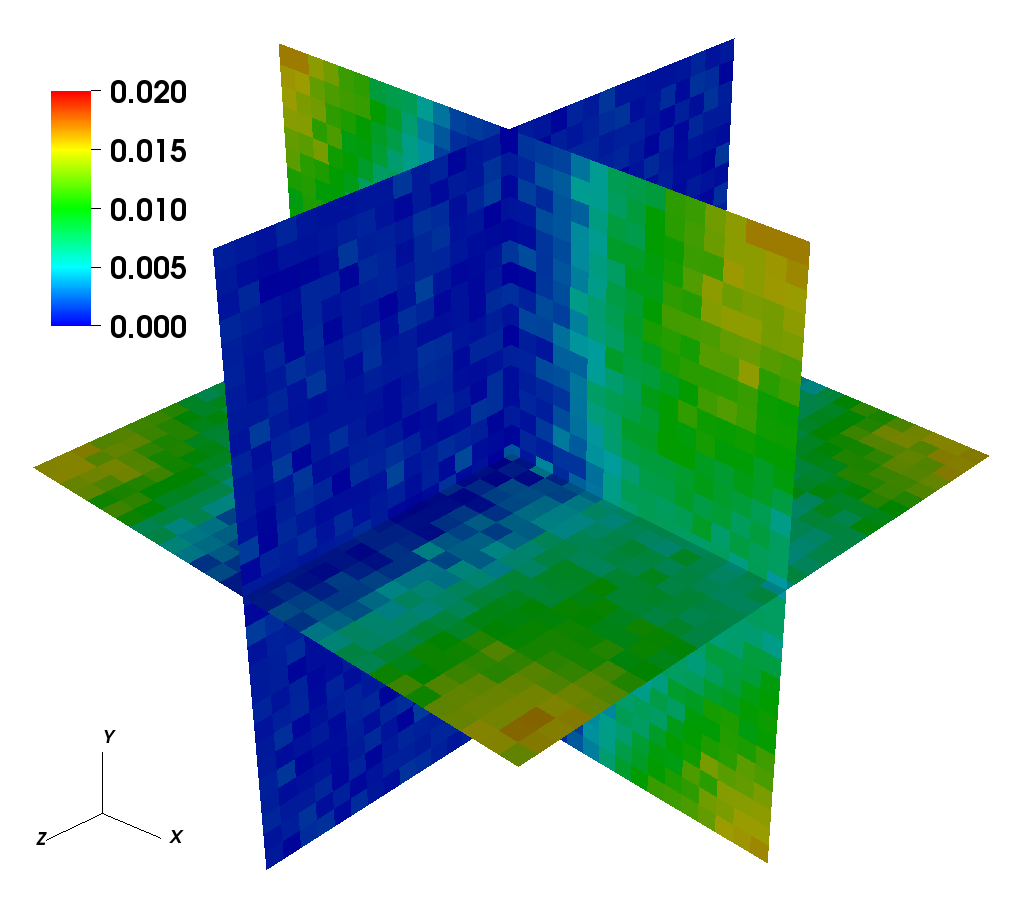}\includegraphics[width=0.45\textwidth]{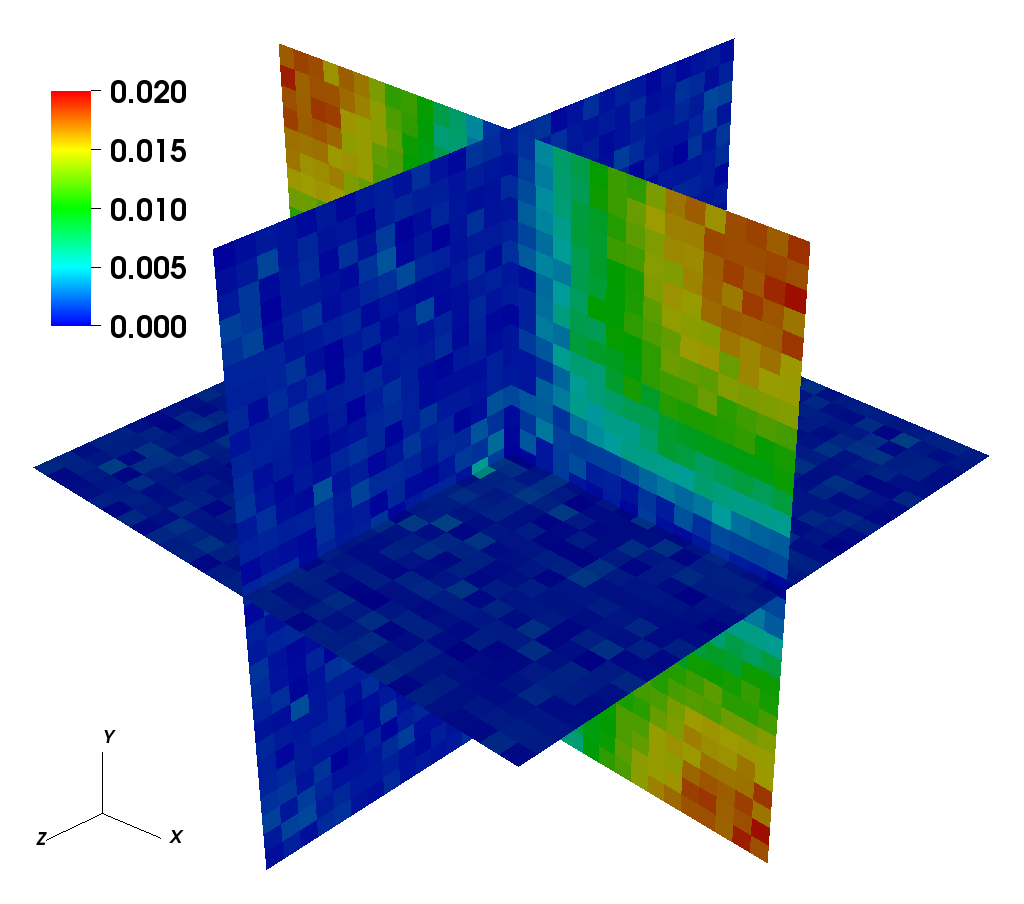}
\par\end{centering}

\caption{\label{fig:S_rho_v_comp}Normalized static structure factors $\tilde{S}_{\rho,\rho}$
(top left), $\tilde{S}_{v_{x},v_{x}}$ (top right), $\tilde{S}_{\rho,v_{x}}$
(bottom left) and $\tilde{S}_{v_{x},v_{y}}$ (bottom right) for a
compressible fluid with physical properties similar to water, for
a periodic system with $30^{3}$ cells. A uniform background flow
with velocity $v_{0}=(0.2,0.1,0.05)c_{T}$ is imposed and the time
step corresponds to an acoustic CFL number $\alpha=0.25$ and viscous
CFL number $\beta_{\nu}=0.017$ for shear viscosity and $\beta_{\zeta}=0.041$
for bulk viscosity.}
\end{figure*}

We first examine the equilibrium discrete Fourier spectra of the density
and velocity fluctuations for a uniform periodic system with an imposed
background flow, with similar results observed for concentration fluctuations.
In Fig. \ref{fig:S_rho_v_comp} we show the correlations of density
and velocity fluctuations as a function of the wavenumber $\V k$
in three dimensions for a CFL number of $\alpha=0.25$. We see that
self-correlations are close to unity while cross-correlations nearly
vanish, as required, with density fluctuations having the largest
relative error of 5\% for the largest wavenumbers. 

Calculating cross-correlations in real space is complicated by the
staggering of the diffent grids. We arbitrarily associate one half
of the cell faces with the cell center, defining $\av{\left(\d{\rho}\right)\left(\d{v_{x}}\right)}\equiv\av{\left(\d{\rho}_{i,j}\right)\left(\d{v_{i+\frac{1}{2},j}^{(x)}}\right)}$
and $\av{\left(\d{v_{x}}\right)\left(\d{v_{y}}\right)}\equiv\av{\left(\d{v_{i+\frac{1}{2},j}^{(x)}}\right)\left(\d{v_{i,j+\frac{1}{2}}^{(y)}}\right)}$.
Theoretical analysis suggests that the error in the discrete covariance
vanishes with time step as $O(\alpha^{3})\sim O\left(c_{T}^{3}\D t^{3}\right)$
\cite{DFDB}. In the right panel of Fig. \ref{fig:error_vs_alpha}
we show the relative error in the discrete covariances as a function
of $\alpha$ in the presence of a background flow, confirming the
predicted cubic convergence. These numerical results verify that our
spatial discretization satisfies discrete fluctuation-dissipation
balance and the temporal discretization is weakly third-order accurate.

\subsubsection{Dynamic Correlations}

For compressible flow, the dynamics of the fluctuations is affected
by the presence of sound waves and it is important to verify that
the numerical scheme is able to reproduce the temporal correlations
between the fluctuations of the different pairs of variables. In particular,
a good method should reproduce the dynamic correlations at small wavenumbers
and wave-frequencies correctly \cite{LLNS_S_k}. Theoretical predictions
for the equilibrium covariances of the spatio-temporal specta of the
fluctuating fields, usually referred to as \emph{dynamic structure
factors}, are easily obtained by solving the equations (\ref{eq:LLNS_rho_simp},\ref{eq:LLNS_comp_v_simp})
in the Fourier wavevector-frequency $(\V k,\omega)$ domain and averaging
over the fluctuations of the stochastic forcing \cite{FluctHydroNonEq_Book}.
The density-density dynamic structure factor $S_{\rho,\rho}(\V k,\omega)$
is accessible experimentally via light scattering measurements, and
for isothermal flow it exhibits two symmetric Brilloin peaks at $\omega\approx\pm c_{T}k$.
The velocity components exhibit an additional central Rayleigh peak
at $\omega=0$ due to the viscous dissipation. As the fluid becomes
less compressible (i.e., the speed of sound increases), there is an
increasing separation of time-scales between the side and central
spectral peaks, showing the familiar numerical stiffness of the compressible
Navier-Stokes equations.

\begin{figure*}
\centering{}\includegraphics[width=0.45\textwidth]{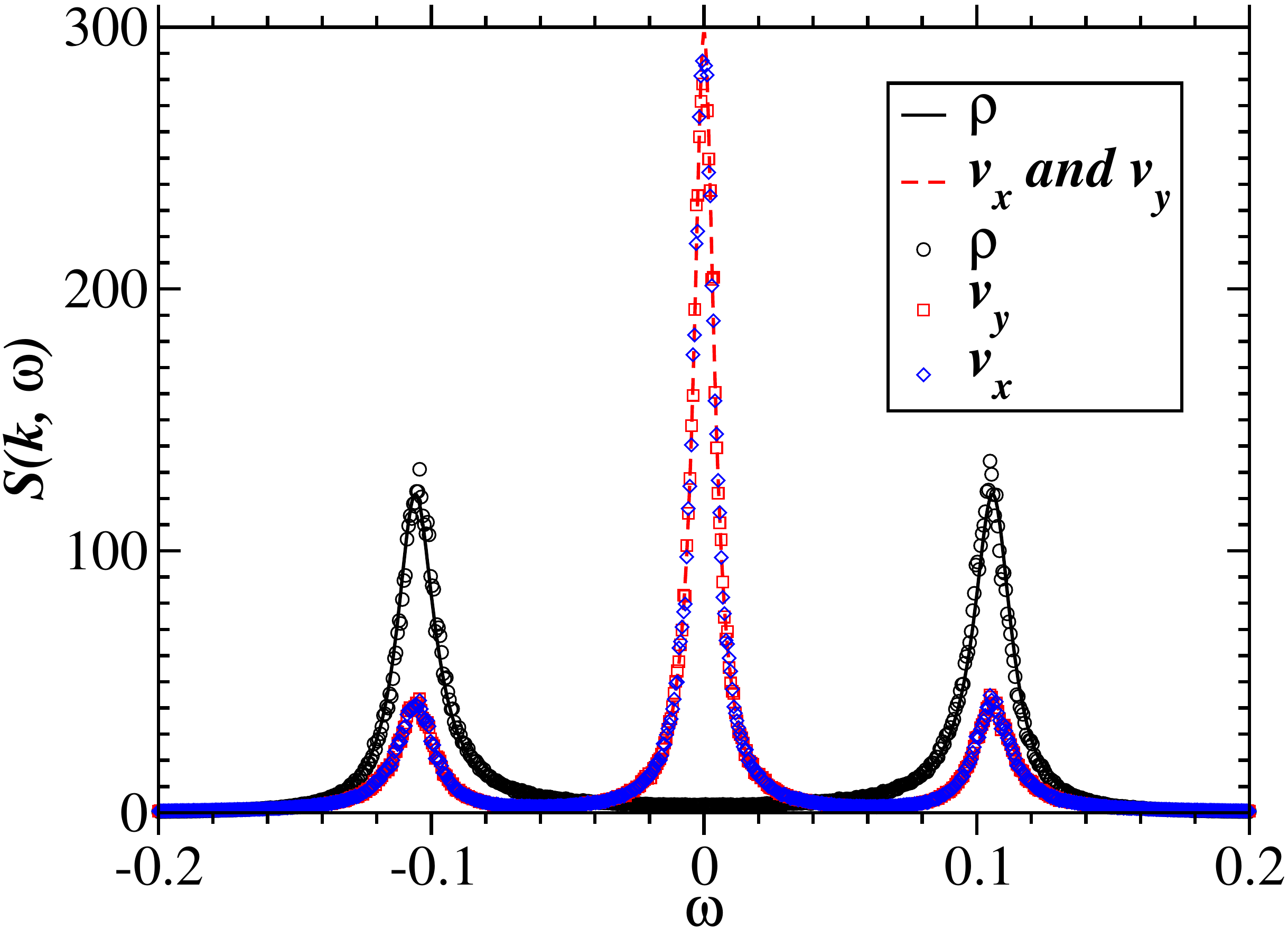}
\includegraphics[width=0.45\textwidth]{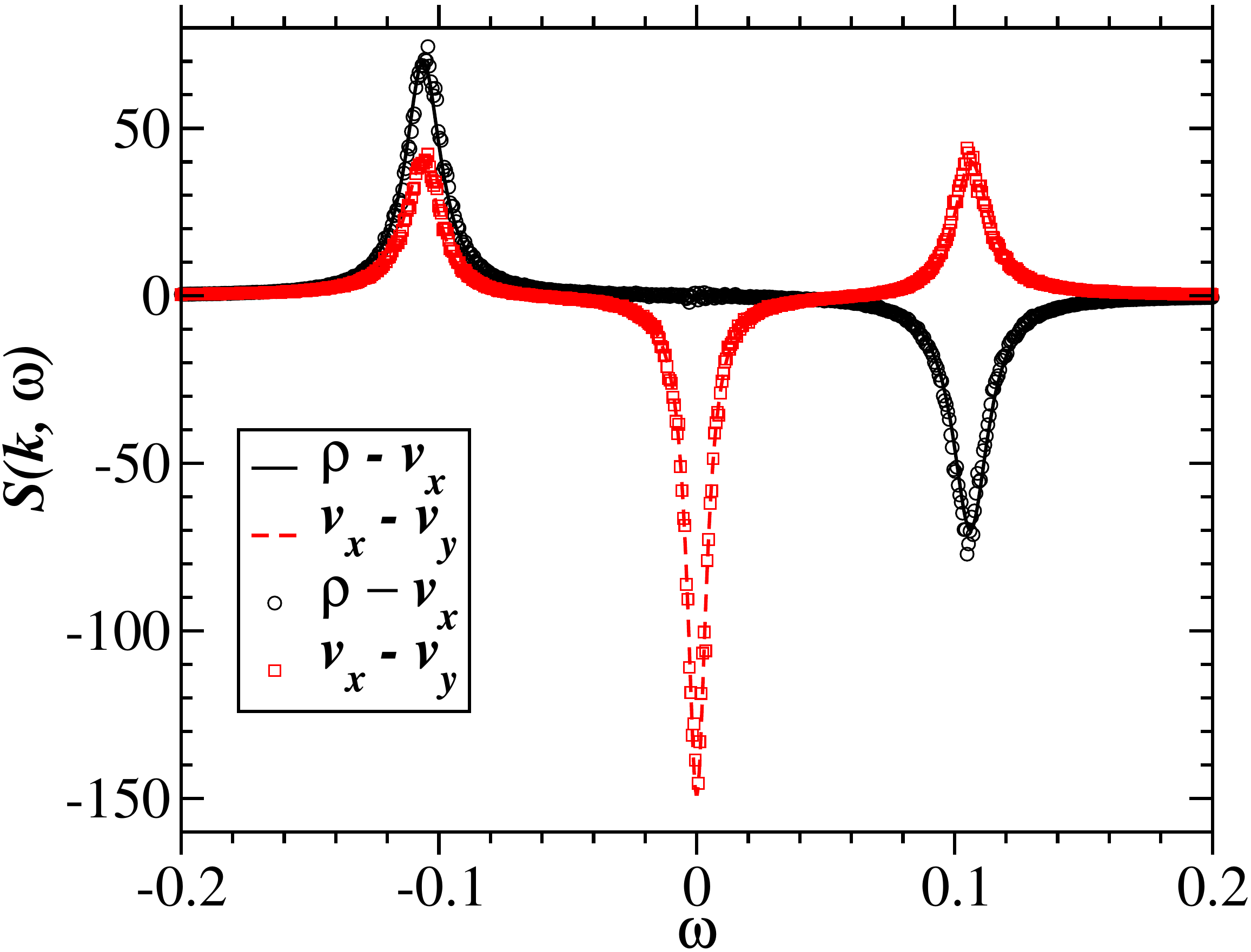}
\caption{\label{fig:S_kw_comp}Numerical data (symbols) and theory (lines)
for the real part of several dynamic structure factors for wavenumber
$\V k=(2,2,2)\cdot2\pi/L$ in a cubic periodic box of $30^{3}$ cells
and volume $L^{3}$. Self correlations are shown in the left panel,
and cross-correlations are shown in the right panel. The imaginary
part vanishes to within statistical accuracy for the off-diagonal
terms. The physical parameters are as reported in the caption of Fig.
\ref{fig:S_rho_v_comp}.}
\end{figure*}

In Fig. \ref{fig:S_kw_comp} we compare the theoretical to the numerical
dynamic structure factors for one of the smallest resolved wavenumbers,
and observe very good agreement. Note that unlike static correlations,
dynamic correlations are subject to discretization artifacts for larger
wavenumbers, even as $\D t\rightarrow0$ \cite{LLNS_S_k}. Specifically,
the positions and widths of the various peaks are set by the effective
wavevector $\tilde{\V k}$ rather than the true wavevector $\V k$,
as given for the standard second-order discretization of diffusion
in (\ref{eq:k_x_tilde}).

\section{\label{sec:Giant-Fluctuations}Giant Fluctuations}

As a non-trivial application of our staggered schemes for fluctuating
hydrodynamics, we perform the first incompressible computer simulations
of diffusive mixing in microgravity, recently studied experimentally
aboard a satellite in orbit around the Earth \cite{FractalDiffusion_Microgravity}.
The experimental data presented in Ref. \cite{FractalDiffusion_Microgravity}
shows good agreement with theoretical predictions, however, various
over-simplifications are made in the theory, notably, only the solenoidal
velocity mode with the largest wavelength is considered. Numerical
simulations allow for a more detailed comparison of experimental data
with fluctuating hydrodynamics, at least within the applicability
of the physical approximations discussed in Section \ref{sub:DetEquations}.

The experimental configuration consists of a dilute solution of polystyrene
in toluene, confined between two parallel transparent plates that
are a distance $h=1\mbox{mm}$ apart. A steady temperature gradient
$\nabla T=\D T/h$ is imposed along the $y$ axes via the plates.
The weak temperature gradient leads to a strong concentration gradient
$\grad\bar{c}=\bar{c}S_{T}\grad T$ due to the Soret effect, giving
rise to an exponential steady-state concentration profile $\bar{c}(y)$.
Quantitative shadowgraphy is used to observe and measure the strength
of the fluctuations in the concentration around $\bar{c}$ via the
change in the refraction index. The observed light intensity, once
corrected for the optical transfer function of the equipment, is proportional
to the intensity of the fluctuations in the concentration averaged
along the gradient, \[
c_{\perp}(x,z)=h^{-1}\int_{y=0}^{h}c(x,y,z)dy.\]
The main physical parameters we employed in our simulations are summarized
in Table \ref{tab:SimParams}. Additional details of the experimental
setup and parameters are given in Ref. \cite{FractalDiffusion_Microgravity}.

\begin{table}
\begin{centering}
\begin{tabular}{|c|c|c|}
\hline 
Parameter & Value & Notes\tabularnewline
\hline
\hline 
$\rho$ & $0.86$ $\mbox{gr/cm}^{3}$ & On average only if compressible\tabularnewline
\hline 
$\chi(\nu+\chi)$ & $1.2\cdot10^{-8}$ $\mbox{cm}^{4}/\mbox{s}^{2}$ & Kept constant in all runs\tabularnewline
\hline 
$\nu$ & Variable $S_{c}=\nu/\chi$ & Physical value $\nu=6.07\cdot10^{-3}\mbox{ cm}^{2}/\mbox{s}$\tabularnewline
\hline 
$\chi$ & Variable $S_{c}=\nu/\chi$ & Physical value $\chi=1.97\cdot10^{-6}\mbox{ cm}^{2}/\mbox{s}$\tabularnewline
\hline 
$\zeta$ & 0 & None for incompressible\tabularnewline
\hline 
$k_{B}T$ & $4.18\cdot10^{-14}\mbox{ gr cm}^{2}/\mbox{s}$ & Corresponds to $T=303\mbox{ K}$\tabularnewline
\hline 
$M$ & $1.51\cdot10^{-20}\mbox{ gr}$ & Not important for results\tabularnewline
\hline 
$S_{T}$ & $0.0649\mbox{ K}^{-1}$ & Enters only via $S_{T}\grad T$\tabularnewline
\hline 
$c_{0}$ & $0.018$ & On average only if nonperiodic\tabularnewline
\hline 
$c_{T}$ & $1.11\mbox{ cm}/\mbox{s}$ & Physical value $c_{T}\approx1.3\cdot10^{5}\mbox{ cm}/\mbox{s}$\tabularnewline
\hline
\end{tabular}
\par\end{centering}

\caption{\label{tab:SimParams}Summary of parameters used in the simulations
of giant fluctuations in zero gravity.}
\end{table}

The large speed of sound in toluene makes the compressible equations
very stiff at the length scales of the experimental system. It is
usually argued that compressibility does not affect the concentration
fluctuations \cite{FluctHydroNonEq_Book}. Solving the compressible
equations in the presence of a concentration gradient confirms that,
as long as there is a large separation of time scales between the
acoustic and diffusive dynamics, the presence of sound waves does
not affect the concentration fluctuations. In our compressible simulations,
we artificially decrease the speed of sound many-fold and set the
cell Reynolds number to $r=c_{T}/(\nu\D x)\geq10$. Numerical results
show that this is sufficient to approach the limit $r\rightarrow\infty$
to within the statistical accuracy of our results. This decrease in
$c_{T}$ corresponds to making the mass of the toluene molecules much
larger than the mass of the polystyrene macromolecules themselves,
which is of course physically very unrealistic. One can think of our
compressible simulations of giant fluctuations in microgravity as
an artificial compressibility method for solving the incompressible
equations.

In the actual experiments reported in Ref. \cite{FractalDiffusion_Microgravity},
concentration diffusion is much slower than momentum diffusion, corresponding
to Schmidt number $S_{c}=\nu/\chi\approx3\cdot10^{3}$. This level
of stiffness makes direct simulation of the temporal dynamics of the
fluctuations infeasible, as long averaging is needed to obtain accurate
steady-state spectra, especially for small wavenumbers. However, as
far as the nonequilibrium static correlations are concerned, we see
from (\ref{eq:S_cc_neq}) that the crucial quantity is $\chi(\nu+\chi)=(s+1)\chi^{2}$,
rather than $\chi$ and $\nu$ individually. Therefore, we can artificially
increase $\chi$ and decrease $\nu$ to reduce $s$, keeping $s\gg1$
and $(s+1)\chi^{2}$ fixed. In the linearized case, it can be proven
more formally that there exists a limiting stochastic process for
the concentration as $s\rightarrow\infty$ so long as $s\chi^{2}$
is kept constant (E. Vanden-Eijnden, private communication). In fact,
artificially decreasing the Schmidt number while keeping $s\chi^{2}$
fixed can be seen as an instance of the \emph{seamless} multiscale
method presented in Ref. \cite{SeamlessMultiscale}.

\subsection{Approximate Theory}

For large wavenumbers the influence of the boundaries can be neglected
and the periodic theory presented in Section \ref{sub:TheoryIncomp}
applied. In order to demonstrate the importance of the boundaries,
and also to test the code by comparing to the periodic theory, we
have implemented a model in which qualitatively similar giant concentration
fluctuations appear even though the macroscopic concentration profile
is uniform, $\bar{c}(y)=c_{0}$. Numerically, this sort of quasi-periodic
model is implemented by using periodic boundary conditions but adding
an additional source term $-\V v\cdot\grad\bar{c}$ in the concentration
equation, as in (\ref{eq:incomp_quasi_periodic}). This term mimics
our skew-adjoint discretization of the advection by the fluctuating
velocities\[
\V v\cdot\grad\bar{c}\rightarrow\left(\M D\M U\bar{\V c}\right)_{i,j}=\frac{\nabla\bar{c}}{2}\left(v_{i,j+\frac{1}{2}}^{(y)}+v_{i,j-\frac{1}{2}}^{(y)}\right),\]
and is conservative when integrated over the whole domain. Note that
in this quasi-periodic setup $\grad\bar{c}$ is simply an externally-imposed
quantity unrelated to the actual mean concentration profile. We emphasize
that these quasi-periodic simulations are used only for testing and
theoretical analysis of the problem, and not for comparison with the
experimental results. In the simulations with physical boundaries
and in the experiments the concentration profile is exponential rather
than linear. For the purposes of constructing a quasi-periodic approximation
we take the effective concentration gradient to be $\nabla\bar{c}\approx\D c/h$,
where $\D c$ is the difference in concentration near the two boundaries.

For periodic systems, the spectrum of the fluctuations of $c_{\perp}$
can be obtained from the full three-dimensional spectrum (\ref{eq:S_cc_neq})
by setting $k_{y}=k_{\parallel}=0$. For the specific parameters in
question the equilibrium fluctuations in concentration are negligible
even at the largest resolved wavenumbers. When discretization artifacts
are taken into account, the quasi-periodic theoretical prediction
for the experimentally-observed spectrum becomes \begin{equation}
S_{\text{QP}}^{\perp}\left(k_{x},k_{z}\right)=\av{\left(\widehat{\d c}_{\perp}\right)\left(\widehat{\d c}_{\perp}\right)^{\star}}=\frac{k_{B}T}{\rho\left[\chi(\nu+\chi)\right]\tilde{k}_{\perp}^{4}}\,\left(\nabla\bar{c}\right)^{2},\label{eq:S_cc_QP}\end{equation}
where $\tilde{k}_{\perp}^{4}=\left(\tilde{k}_{x}^{2}+\tilde{k}_{z}^{2}\right)^{2}$
and tilde denotes the effective wavenumber (\ref{eq:k_x_tilde}).
Imposing no-slip conditions for the fluctuating velocities makes the
theory substantially more complicated. A single-mode approximation
for the velocities is made in Ref. \cite{RayleighBernard_LLNS} in
order to obtain a closed-form expression for the spectrum of concentration
fluctuations in a non-periodic system $S_{\text{NP}}^{\perp}$. For
a small Lewis number and without gravity it is found that \begin{equation}
\frac{S_{\text{NP}}^{\perp}(\V k_{\perp})}{S_{\text{QP}}^{\perp}(\V k_{\perp})}\approx G(hk_{\perp})=\frac{q_{\perp}^{4}}{q_{\perp}^{4}+24.6q_{\perp}^{2}+500.5},\label{eq:Galerkin_function}\end{equation}
where $q_{\perp}=hk_{\perp}$ is a non-dimensionalized wavenumber.

\begin{figure}
\begin{centering}
\includegraphics[width=0.3\textwidth]{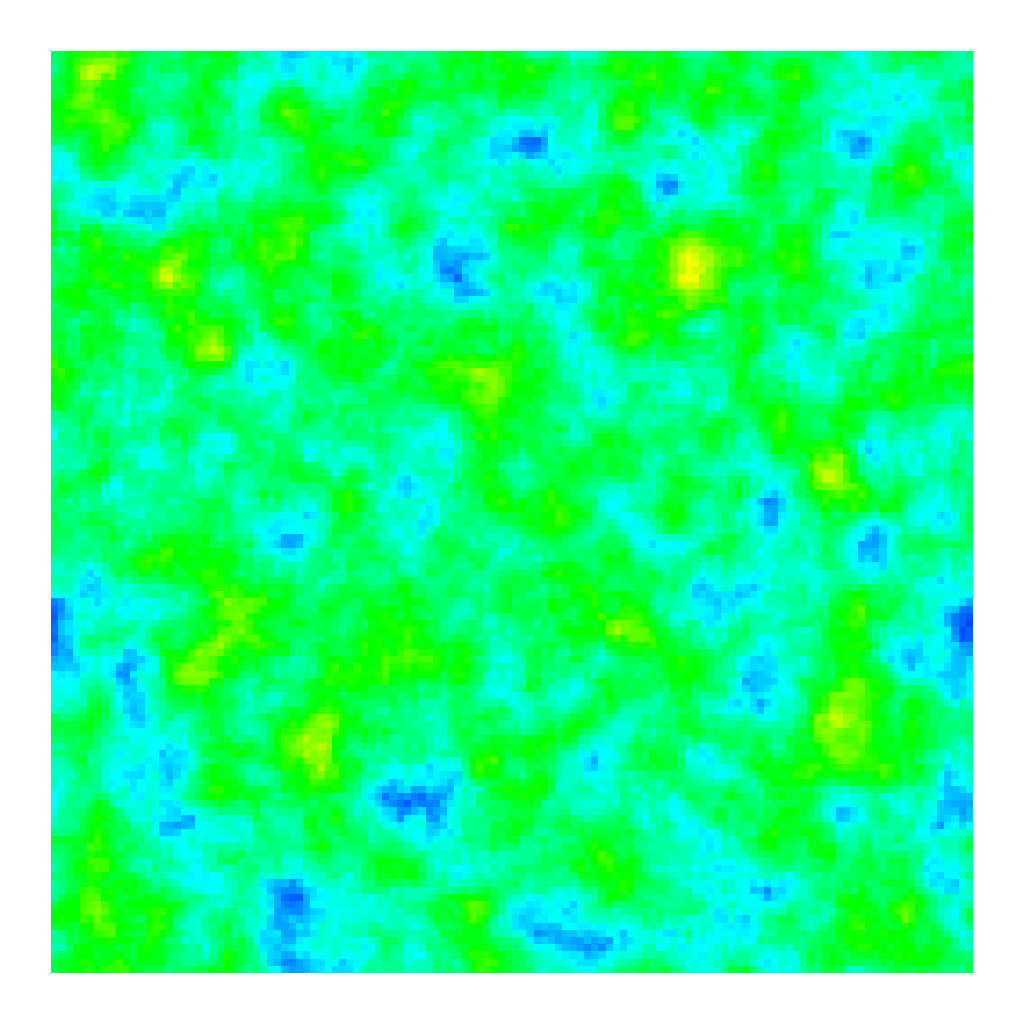}\includegraphics[width=0.3\textwidth]{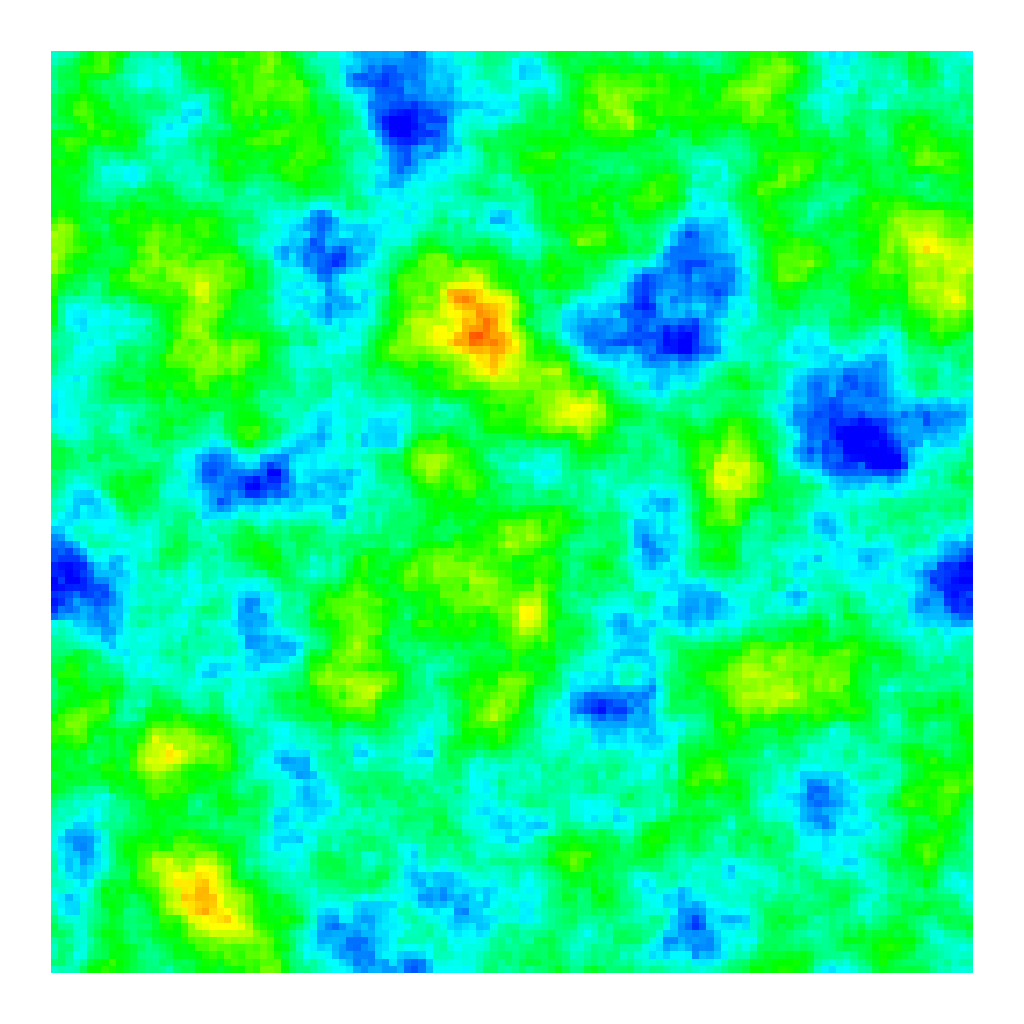}\includegraphics[width=0.3\textwidth]{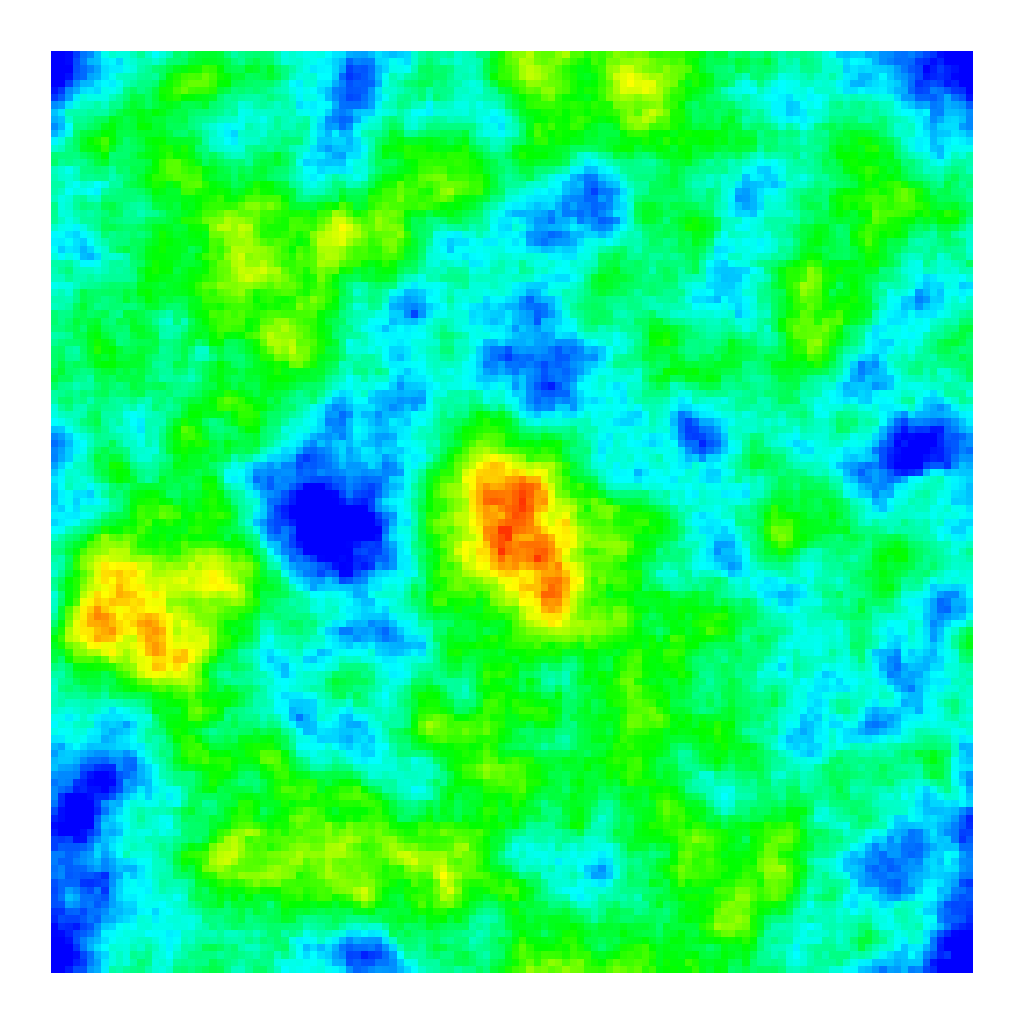}
\par\end{centering}

\caption{\label{fig:c_perp_snapshots}Snapshots of the concentration $c_{\perp}$
in the plane perpendicular to the gradient $\grad\bar{c}$, at times
$0.1\tau_{0}$, $\tau_{0}$, and $5\tau_{0}$ after the gradient is
established. The thickness of the sample (perpendicular to the page)
is one quarter of the lateral extents of the system, $h=L_{y}=L_{x}/4$,
and sets the scale of the steady-state fluctuations. Compare to the
experimental snapshots shown in Fig. 1 of Ref. \cite{FractalDiffusion_Microgravity}.}
\end{figure}

The Galerkin function $G$ given by (\ref{eq:Galerkin_function})
reflects the physical intuition that the no-slip condition suppresses
fluctuations at scales larger than the distance between the physical
boundaries \cite{FractalDiffusion_Microgravity}. After the concentration
gradient is established, {}``giant'' \cite{GiantFluctuations_Nature}
concentration fluctuations evolve with a typical time scale of $\tau_{0}=h^{2}/(\pi^{2}\chi)\sim1000\mbox{s}$,
until a steady state is reached in which the typical length scale
of the concentration fluctuations is set by the finite extent of the
domain. This is illustrated in Fig. \ref{fig:c_perp_snapshots} via
snapshots of $c_{\perp}(x,z;\, t)$ taken at several points in time
after starting with no concentration fluctuations at time $t=0$.

\subsection{Simulations and Results}

In our simulations, the plates are represented by no-slip boundaries
at $y=0$ and $y=h$, and periodic boundaries are imposed along the
$x$ and $z$ axis to mimic the large extents of the system in the
directions perpendicular to the gradient. A Robin boundary condition
is used for concentration at the physical boundary,\[
\frac{\partial c}{\partial n}=-c\left(\V n\cdot\V v_{s}\right),\]
ensuring that the normal component of the concentration flux vanishes
at a physical boundary. The stochastic concentration flux also vanishes
at the boundary as for Dirichlet boundaries, since the Soret term
does not affect fluctuation-dissipation balance. In the codes the
boundary condition is imposed by setting the concentration in a ghost
cell to \[
c_{g}=c_{n}\frac{2\pm v_{s}\D y}{2\mp v_{s}\D y},\]
where $c_{n}$ is the value in the neighboring cell in the interior
of the computational domain, and the sign depends on whether the ghost
cell is at the low or high end of the $y$ axes. The boundary condition
is imposed explicitly, which leads to non-conservation of the total
concentration when a semi-implicit method is used for the diffusive
terms in the concentration equation. This can be corrected by implementing
the boundary condition implicitly or using an explicit method for
concentration; however, we do not do either since the observed change
in the average concentration is small for the specific parameters
we use.

Using the incompressible formulation allows for a much larger time
step, not only because of the lack of acoustics, but also because
of the implicit temporal discretization of the viscous terms in the
momentum equations. However, it is important to remember that a time
step of our GPU-parallelized compressible code takes much less computing
than a time step of the incompressible code. Nevertheless, we are
able to study larger system sizes in three dimensions using the incompressible
algorithm. In the incompressible sijmulations, we used (\ref{eq:split_v})
for the velocity equation in order to avoid unnecessary projections.
Because of the explicit handling of the concentration boundary conditions,
we employed a predictor-corrector algorithm for the concentration
equation, in which both the predictor and the corrector stages have
the form (\ref{eq:split_c}).

\begin{figure}
\centering{}\includegraphics[width=0.5\textwidth]{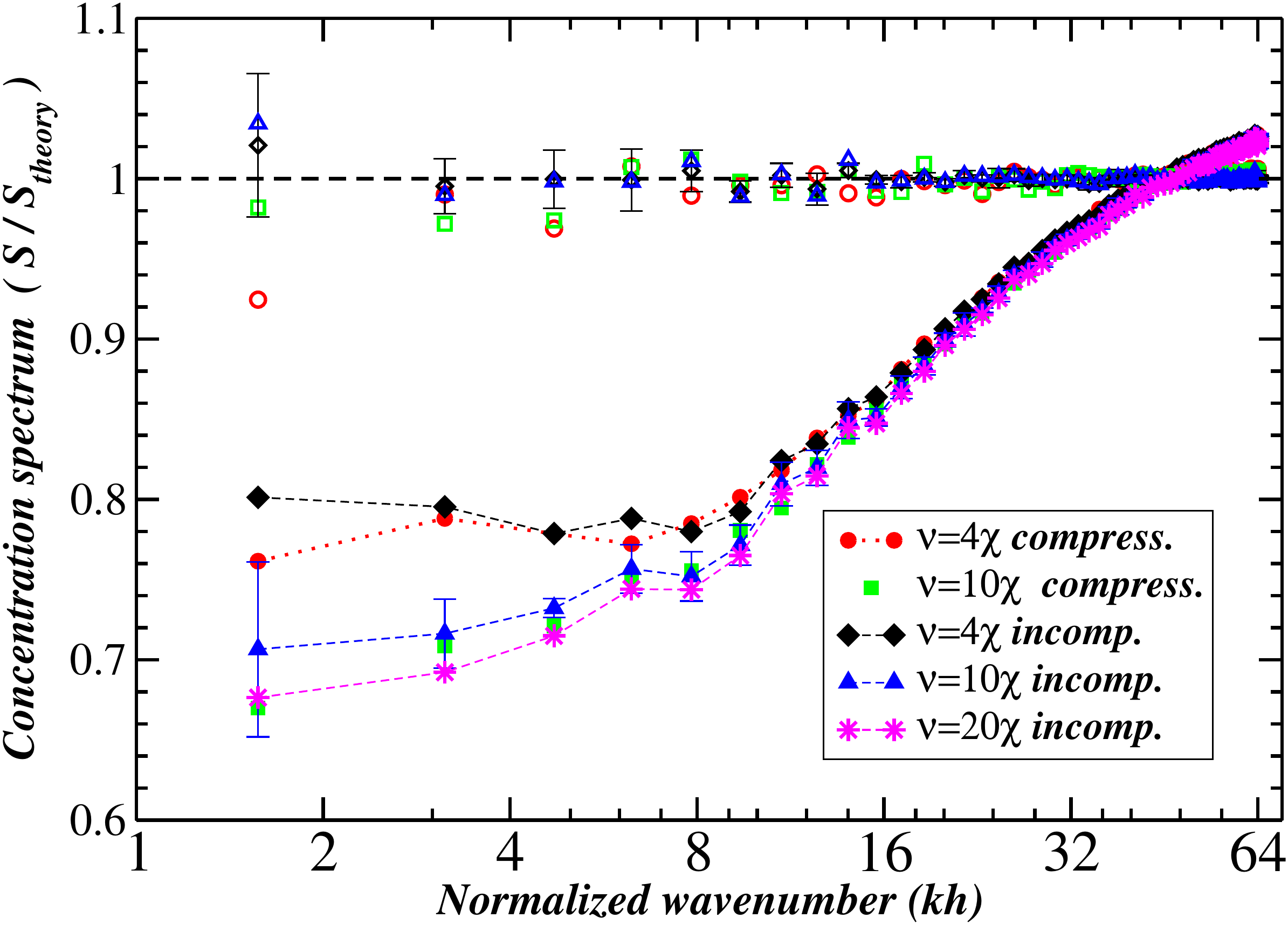}\includegraphics[width=0.5\textwidth]{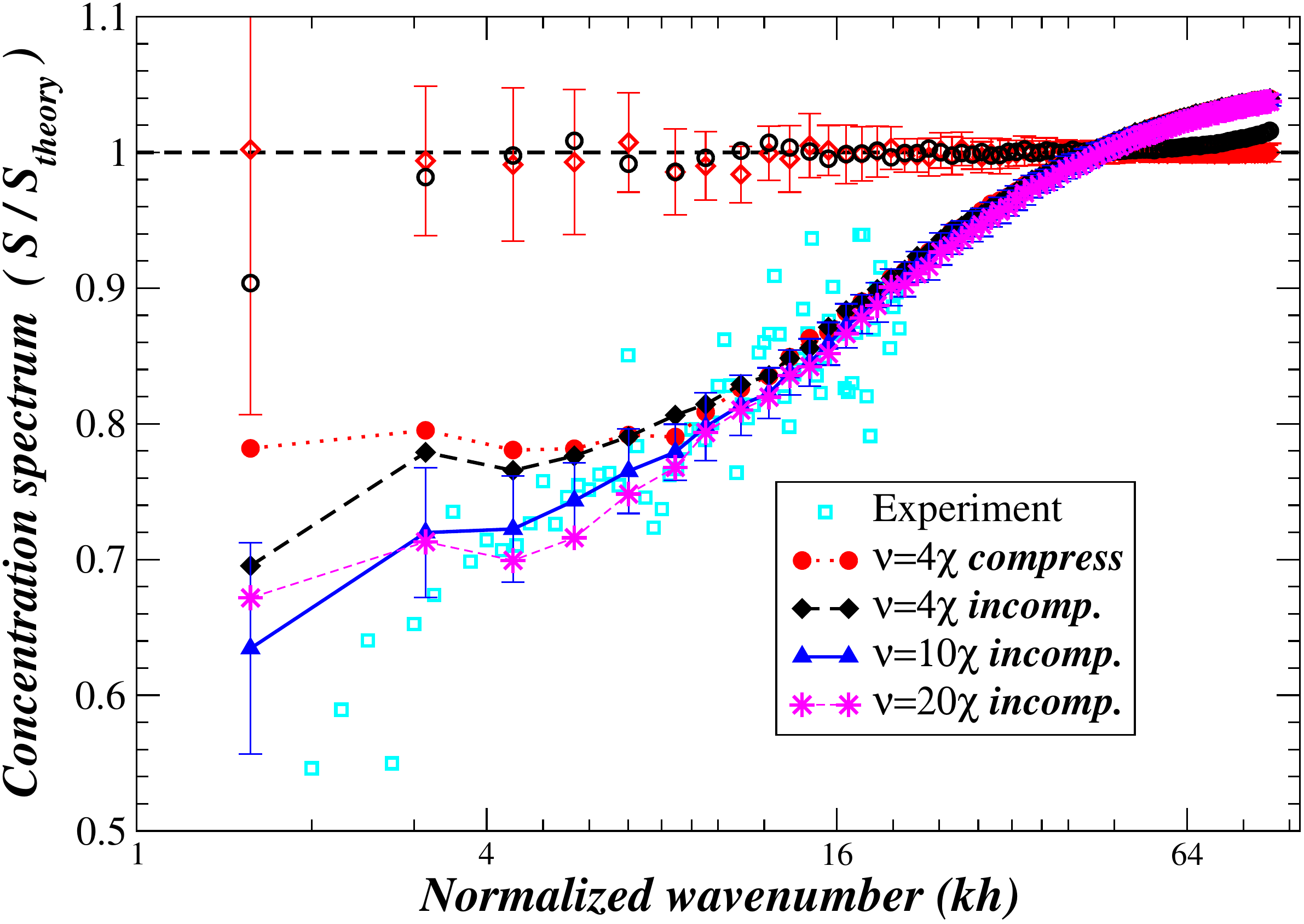}
\caption{\label{fig:S_cc_proj}Ratio between the numerical and theoretical
discrete spectrum of concentration projected along the $y$ axes,
as a function of the normalized wavenumber $q_{\perp}=k_{\perp}h$.
For all runs $N_{y}=32$ cubic hydrodynamic cells along the $y$ axes
were used, and all systems have aspect ratio $N_{x}/N_{y}=N_{z}/N_{y}=4$.
Error bars are indicated for some of the curves to indicate the typical
level of statistical uncertainty. (\emph{Left}) Two dimensions, for
both compressible and incompressible fluids (see legend), with either
periodic boundary conditions (empty symbols) or physical boundaries
(solid symbols) imposed at the $y$-boundaries, for several Schmidt
numbers $s=\nu/\chi$. (\emph{Right}) Three dimensions, with same
symbols as left panel), along with arbitrarily normalized experimental
data \cite{FractalDiffusion_Microgravity} (see legend) corresponding
to $s\approx3\cdot10^{3}$ (courtesy of A. Vailati).}
\end{figure}

In Fig. \ref{fig:S_cc_proj} we show numerical results for the steady-state
spectrum of the discrete concentration field averaged along the $y$-axes,
in two (left panel) and in three dimensions (right panel), for both
bulk (quasi-periodic) and finite (non-periodic) systems. In order
to compare with the theoretical predictions (\ref{eq:S_cc_QP}) and
(\ref{eq:Galerkin_function}) most directly, we plot the ratio of
the observed to the predicted spectrum. This choice of normalization
not only emphasizes any mismatch with the theory, but also eliminates
the power-law ($k_{\perp}^{-4}$) divergence and makes it easier to
average over nearby wavenumbers $\tilde{k}_{\perp}$ and also estimate
error bars%
\footnote{Note, however, that the most reliable error bars are obtained by averaging
over many \emph{uncorrelated }runs started with different random number
seeds.%
}. For the runs reported in Fig. \ref{fig:S_cc_proj} we applied the
largest concentration (temperature) gradient ($\D T=17.4\mbox{K}$)
used in the experiments \cite{FractalDiffusion_Microgravity}; we
have verified that the non-equilibrium concentration fluctuations
scale as the square of the gradient.

Both panels in Fig. \ref{fig:S_cc_proj} show an excellent agreement
between the theory (\ref{eq:S_cc_QP}) and the numerical results for
quasi-periodic systems. This shows that correcting for the spatial
discretization artifacts by replacing $k_{\perp}$ with $\tilde{k}_{\perp}$
accounts for most of the discretization error. For the compressible
runs, we use a relatively small time step, $\alpha=0.2$, leading
to temporal discretization errors that are smaller than the statistical
accuracy except at the largest wavenumbers. Our semi-implicit discretization
of the incompressible equations gives the correct static covariance
of the concentration for \emph{all} time step sizes. Based on the
analysis presented in Appendix \ref{sec:CNAccuracy}, the majority
of the incompressible simulations employ a time step corresponding
to a viscous CFL number $\beta=1$ or $\beta=2$, with a few of the
largest systems run at $\beta=5$ to resolve the smaller wavenumbers
better. 

In the left panel of Fig. \ref{fig:S_cc_proj} we compare results
from two-dimensional compressible and incompressible simulations and
find excellent agreement. For non-periodic systems the single-mode
Galerkin theory (\ref{eq:Galerkin_function}) is not exact and the
theory visibly over-predicts the concentration fluctuations for smaller
wavenumbers in both two and three dimensions. We observe only a partial
overlap of the data for different Schmidt numbers $s=\nu/\chi$ for
smaller wavenumbers, although the difference between $s=10$ and $s=20$
is relatively small.

In three dimensions, we rely on the incompressible code in order to
reach time scales necessary to obtain sufficiently accurate steady-state
averages for large Schmidt numbers. In the right panel of Fig. \ref{fig:S_cc_proj}
we compare numerical results for quasi-periodic and non-periodic compressible
and incompressible systems to the theoretical predictions and also
to experimental data from Ref. \cite{FractalDiffusion_Microgravity}
(A. Vailati, private communication). While the numerical data do not
match the experiments precisely at the smallest wavenumbers, a more
careful comparison is at present not possible. Firstly, the boundary
conditions affect the small wavenumbers strongly, and our use of periodic
boundary conditions in $x$ and $z$ directions does not match the
experimental setup. The experimental data has substantial measurement
uncertainties, and is presently normalized by an arbitrary pre-factor.
Within this arbitrary normalization, our numerical results seem to
be in good agreement with the experimental observations over the whole
range of experimentally-accessible wavenumbers, and the agreement
at small wavenumbers improves as the Schmidt number of the simulations
increases. The actual magnitude of the \emph{macroscopic} non-equilibrium
fluctuations in $c_{\perp}$ is given by the integral of the structure
factor $S_{c,c}^{\perp}$ over all wavenumbers $\V k_{\perp}$. Numerically
we observe fluctuations $\av{\left(\d c_{\perp}\right)^{2}}/\bar{c}_{\perp}^{2}\approx3\cdot10^{-7}$,
which is consistent with experimental estimates (A. Vailati, private
communication).

\section{Conclusions}

We have presented spatio-temporal discretizations of the equations
of fluctuating hydrodynamics for both compressible and incompressible
mixtures of dynamically-identical isothermal fluids. As proposed by
some of us in Ref. \cite{LLNS_S_k}, we judge the weak accuracy of
the schemes by their ability to reproduce the equilibrium covariances
of the fluctuating variables. In particular, for small time steps
the spatial discretization ensures that each mode is equally forced
and dissipated in agreement with the fluctuation-dissipation balance
principle satisfied by the continuum equations. A crucial ingredient
of this discrete fluctuation-dissipation balance is the use of a discrete
Laplacian $\M L=-\M D\M D^{\star}$ for the dissipative fluxes, where
$\M D$ is a conservative discrete divergence, with a suitable correction
to both the Laplacian stencil and the stochastic fluxes at physical
boundaries. Furthermore, we utilize a centered skew-adjoint discretization
of advection which does not additionally dissipate or force the fluctuations,
as previously employed in long-time simulations of turbulent flow,
where it is also crucial to ensure conservation and avoid artificial
dissipation \cite{ConservativeDifferences_Incompressible}.

For the compressible equations, our spatio-temporal discretization
is closely based on the collocated scheme proposed by some of us in
Ref. \cite{LLNS_S_k}, except that here we employ a staggered velocity
grid. It is important to point that out the difference between a \emph{collocated}
scheme, in which the fluid variables are cell-centered but the stochastic
fluxes are face-centered (staggered), as described in Ref. \cite{LLNS_S_k},
and a \emph{centered} scheme where \emph{all} quantities are cell-centered.
Several authors \cite{Delgado:08,StagerredFluctHydro} have already
noted that centered schemes lead to a Laplacian that decouples neighboring
cells, which is problematic in the context of fluctuating hydrodynamics.
We emphasize however that these problems are not shared by collocated
schemes for compressible fluids, for which the Laplacian $\M L=-\M D\M D^{\star}$
has the usual compact $2d+1$ stencil, where $d$ is the dimensionality
\cite{LLNS_S_k}. Discretizations in which all conserved quantities
are collocated may be preferred over staggered ones in particle-continuum
hybrids \cite{DSMC_Hybrid}, or more generally, in conservative discretizations
for non-uniform grids.

A staggered grid arrangement, however, has a distinct advantage for
incompressible flow. Namely, the use of a staggered grid simplifies
the construction of a robust idempotent discrete projection $\M{\Set P}=\M I+\M D^{\star}\M L^{-1}\M D$
that maintains discrete fluctuation-dissipation at all wavenumbers.
In the temporal discretization employed here, based on prior work
by one of us \cite{NonProjection_Griffith}, this projection is used
as a preconditioner for solving the Stokes equations for the pressure
and velocities at the next time step. For periodic systems the method
becomes equivalent to a classical Crank-Nicolson-based projection
method, while at the same time avoiding the appearance of artificial
pressure modes in the presence of physical boundaries \cite{ProjectionModes_III,GaugeIncompressible_E}.

The numerical results presented in Section \ref{sec:Giant-Fluctuations}
verify that our numerical simulations model experimental measurements
of giant fluctuations \cite{FractalDiffusion_Microgravity} during
diffusive mixing of fluids faithfully. The numerical simulations give
access to a lot more data than experimentally measurable. For example,
the spectrum of concentration fluctuations in the $x-z$ plane can
be computed for planes (slices) as the distance from the boundaries
is varied, giving a more complete picture of the three dimensional
spatial correlations of the nonequilibrium fluctuations. We defer
a more detailed analysis, including a study of temporal correlations,
to future work.

The compressible solver we developed utilizes modern GPUs for accelerating
the computations. In the future we will investigate the use of GPUs
for the incompressible equations, starting with simple FFT-based solvers
for periodic systems. For grid sizes that are much larger than molecular
scales, the stability restriction of explicit compressible solvers
becomes severe and it becomes necessary to eliminate sound waves from
the equations by employing the low Mach number limit. A challenge
that remains to be addressed in future work is the design of zero
Mach number methods \cite{LowMachMinicourse} for solving the variable-density
equations of fluctuating hydrodynamics, as necessary when modeling
mixtures of miscible fluids with different densities. This would enable
computational modeling of the effects of buoyancy (gravity) in experimental
studies of the giant fluctuation phenomenon performed on Earth \cite{GiantFluctuations_Nature,GiantFluctuations_Theory,GiantFluctuations_Universal}.

\section*{Acknowledgments}

We thank Alberto Vailati for insightful comments and sharing experimental
data from the GRADFLEX experiments \cite{FractalDiffusion_Microgravity}.
We thank Alejandro Garcia for a careful reading and suggestions on
improving this work. We thank Eric Vanden-Eijnden and Jonathan Goodman
for numerous inspiring discussions and motivating the Metropolis-Hastings
Monte Carlo argument presented in Appendix \ref{sec:CNAccuracy}.
B. Griffith acknowledges research support from the National Science
Foundation under awards OCI 1047734 and DMS 1016554. J. Bell and A.
Donev were supported by the DOE Applied Mathematics Program of the
DOE Office of Advanced Scientific Computing Research under the U.S.
Department of Energy under contract No. DE-AC02-05CH11231. Additional
support for A. Donev was provided by the National Science Foundation
under grant DMS-1115341. T. Fai wishes to acknowledge the support
of the DOE Computational Science Graduate Fellowship, under grant
number DE-FG02-97ER25308. R. Delgado-Buscalioni and F. Balboa acknowledge
funding from the Spanish government FIS2010-22047-C0S and from the
Comunidad de Madrid MODELICO-CM (S2009/ESP-1691).

\begin{appendix}

\section{\label{sec:CNAccuracy}Implicit Midpoint Rule as a Gibbs Sampler}

We consider here numerical methods for the general additive-noise
linear SDE\begin{equation}
\frac{d\V x}{dt}=\V A\V x+\M K\M{\mathcal{W}}\left(t\right),\label{eq:linear_SDE}\end{equation}
where $\M{\mathcal{W}}(t)$ denotes white noise. If the eigenvalues
of $\M A$ have negative real parts, the long-time dynamics tends
to a Gaussian equilibrium distribution \begin{equation}
P_{\text{eq}}\left(\V x\right)=Z^{-1}\exp\left(-\frac{\V x^{\star}\M S^{-1}\M x}{2}\right),\label{eq:Gibbs_linear}\end{equation}
where the covariance matrix $\M S$ is the solution to the linear
system {[}see, for example, Eq. (30) in \cite{LLNS_S_k} or Eq. (3.10)
in \cite{AMR_ReactionDiffusion_Atzberger}{]}\begin{equation}
\M A\M S+\M S\M A^{\star}=-\M K\M K^{\star}.\label{eq:DFDB_linear}\end{equation}
If one is only interested in calculating steady-state observables
(expectation values), then a numerical method for solving (\ref{eq:linear_SDE})
needs to only sample the equilibrium Gibbs distribution (\ref{eq:Gibbs_linear}),
without having to approximate the correct dynamics.

The implicit midpoint rule or Crank-Nicolson discretization that we
employed in Section \ref{sub:IncompressibleTemporal},\begin{equation}
\V x^{n+1}=\V x^{n}+\M A\left(\frac{\V x^{n}+\V x^{n+1}}{2}\right)\D t+\D t^{1/2}\M K\V W^{n},\label{eq:CN_linear}\end{equation}
can be seen as a Markov Chain Monte Carlo (MCMC) algorithm for sampling
from the distribution (\ref{eq:Gibbs_linear}). This sampling is exact,
that is the equilibrium distribution of the chain (\ref{eq:CN_linear})
is exactly (\ref{eq:Gibbs_linear}). This important fact can be shown
using the techniques described in Ref. \cite{LLNS_S_k}, but here
we present an alternative derivation.

A well-known MCMC algorithm for sampling the Gibbs distribution is
the Metropolis-Hastings algorithm. In this algorithm, one treats $\V x^{n+1}$
as a \emph{trial} or \emph{proposal} move that is then to be accepted
with probability\[
\alpha=\frac{P_{\text{eq}}\left(\V x^{n+1}\right)}{P_{\text{eq}}\left(\V x^{n}\right)}\frac{P_{\text{rev}}\left(\V x^{n+1}\rightarrow\V x^{n}\right)}{P_{\text{forw}}\left(\V x^{n}\rightarrow\V x^{n+1}\right)},\]
where $P_{\text{forw}}$ is the transition probability for the chain
(\ref{eq:CN_linear}) and $P_{\text{rev}}$ is the transition probability
for the \emph{time-reversed} chain (this important distinction ensures
strict time reversibility of the chain with respect to the equilibrium
distribution). Explicitly,\begin{eqnarray*}
P_{\text{rev}}\left(\V x^{n+1}\rightarrow\V x^{n}\right) & = & C\,\exp\left[-\frac{\left(\V W^{n}\right)^{\star}\left(\V W^{n}\right)}{2}\right],\\
P_{\text{forw}}\left(\V x^{n}\rightarrow\V x^{n+1}\right) & = & C\,\exp\left[-\frac{\left(\widetilde{\V W}^{n}\right)^{\star}\left(\widetilde{\V W}^{n}\right)}{2}\right],\end{eqnarray*}
where the reverse step noise $\widetilde{\V W}^{n}$ is the solution
to the equation (here the adjoint of $\M A$ appears because of time
reversal)\[
\V x^{n}=\V x^{n+1}+\M A^{\star}\left(\frac{\V x^{n}+\V x^{n+1}}{2}\right)\D t+\D t^{1/2}\M K\widetilde{\V W}^{n}.\]
A tedious but straightforward matrix calculation shows that the acceptance
probability $\alpha=1$, that is, no rejection is necessary for the
implicit midpoint rule to sample the correct equilibrium distribution,
\emph{regardless} of the time step $\D t$.

The calculation of $\alpha$ is simple to do if a Fourier transform
is used to diagonalize the hydrodynamic equations {[}see Eq. (\ref{eq:c_v_Fourier}){]}
to obtain a system of scalar SDEs with complex coefficients. For the
stochastic advection-diffusion equation (\ref{eq:conservative_formulation})
with $\V v=v_{0}$, which is a good model for more general hydrodynamic
equations,\begin{equation}
\M A\equiv A=-a+b\, i,\quad\M K\equiv K=\sqrt{2a},\mbox{ and }\M S\equiv S=1,\label{eq:scalar_SDE}\end{equation}
with $a=\chi k^{2}$ and $b=-kv_{0}$, where $k$ is the wavenumber.

While the time step $\D t$ can be chosen arbitrarily without biasing
the sampling, the optimal choice is the one that minimizes the variance
of the Monte Carlo estimate of the observable of interest. In the
simulations of giant fluctuation experiments, the observable of interest
is the covariance (spectrum) of the fluctuations $\M S=\av{\M x\M x^{\star}}.$
The variance of the Monte Carlo estimate of $\M S$ is proportional
to the autocorrelation time $\tau$ of $\V S^{n}=\M x^{n}\left(\M x^{n}\right)^{\star}$,
which is itself proportional to the sum of the autocorrelation function
of $\V S^{n}$ \cite{MC_Sokal}. Focusing on the scalar SODE \ref{eq:scalar_SDE},
we get the autocorrelation time\[
\tau\sim\sum_{n=0}^{\infty}\left[\av{S^{k}S^{k+n}}-\av{S^{k}}^{2}\right]=\sum_{n=0}^{\infty}\left(AA^{\star}\right)^{n}=\frac{1}{2a\D t}+\frac{1}{2}+\frac{a\D t}{8}+\frac{b\D t}{a}.\]
For the purely diffusive equation, $v_{0}=0$, the statistical accuracy
for a fixed number of time steps is proportional to

\[
\tau^{-1}=\frac{8\tilde{k}^{2}\beta}{4+4\tilde{k}^{2}\beta+\tilde{k}^{4}\beta^{2}},\]
where $\beta=\nu\D t/\D x^{2}$ is the viscous CFL number and $\tilde{k}=k\D x$
is the dimensionless wavenumber. Note that $\tau^{-1}\sim\beta\tilde{k}^{2}$
for small $\tilde{k}$, so increasing the time step improves the sampling.
However, for large $\tilde{k}$ increasing the time step reduces the
statistical accuracy (this is related to the fact that the Crank-Nicolson
algorithm is $A$-stable but it is \emph{not} $L$-stable), $\tau^{-1}\sim\left(\beta\tilde{k}^{2}\right)^{-1}$.
The wavenumber with highest statistical accuracy $\tilde{k}_{\text{opt}}$
depends on the time step, $\beta\tilde{k}_{\text{opt}}^{2}=2$, or,
alternatively, the optimal choice of time step depends on the wavenumber
of most interest. For the type of problems we studied in this work
the spectrum of the fluctuations has power-law tails $\sim k^{-4}$
and therefore all wavenumbers are important. Using $\beta\sim2$ produces
a good coverage of all of the wavenumbers.

\section{\label{sec:ProjectionAccuracy}Fluctuation-Dissipation Balance for
Incompressible Flow}

Discrete fluctuation-dissipation balance is affected by the presence
of an incompressibility constraint. The spatially discretized velocity
equation linearized around a stationary equilibrium state has the
form, omitting unimportant constants in the noise amplitude,\begin{equation}
\partial_{t}\V v=\M{\Set P}\left[\nu\M L\V v+\sqrt{2\nu}\M D\V W_{\V v}\right],\label{eq:v_t_incomp}\end{equation}
where we used a non-symmetric stochastic stress tensor since the symmetry
does not affect the results presented here. The steady-state covariance
of the velocities $\M S_{\V v}=\left\langle \V v\V v^{\star}\right\rangle $
is determined from the fluctuation-dissipation balance condition (\ref{eq:DFDB_linear})
with $\M A=\nu\M{\Set P}\M L$ and $\M K=\sqrt{2\nu}\,\M{\Set P}\M D$,
giving \begin{equation}
\M{\Set P}\M L\M S_{\V v}+\M S_{\V v}\M L^{\star}\M{\Set P}^{\star}=-2\M{\Set P}\M D\M D^{\star}\M{\Set P}^{\star}.\label{eq:DFDB_incomp}\end{equation}
The fluctuation-dissipation balance condition for the simple advection-diffusion
equation,\[
\M L+\M L^{\star}=\M D\M G+\left(\M D\M G\right)^{\star}=-2\M D\M D^{\star},\]
implies that $\M S_{\V v}=\M{\Set P}$ is the solution to (\ref{eq:DFDB_incomp})
if $\M{\Set P}$ is self-adjoint, $\M{\Set P}^{\star}=\M{\Set P}$,
as stated in (\ref{eq:C_vv_nodt}) with all of the constants included.

The above analysis was does not account for the temporal discretization.
For small timesteps, our temporal discretization of (\ref{eq:v_t_incomp})
behaves like a projected Euler-Maruyama method,\[
\V v^{n+1}=\M{\Set P}\left[\V v^{n}+\nu\M L\V v\D t+\sqrt{2\nu\D t}\M D\V W_{\V v}\right].\]
An important difference with the continuum equation (\ref{eq:v_t_incomp})
is that the velocity in the previous time step is also projected,
i.e., the increment of $O\left(\D t\right)$ is added to $\M{\Set P}\V v^{n}$
and \emph{not} to $\V v^{n}$. If $\M{\Set P}$ is idempotent, $\M{\Set P}^{2}=\M{\Set P}$,
just as the continuum projection operator is, then subsequent applications
of the projection operator do not matter since $\V v^{n}$ is already
discretely divergence free, $\M{\Set P}\V v^{n}=\V v^{n}$. In the
literature on projection methods idempotent projections are called
\emph{exact} projections.

The above considerations lead to the conclusion that $\M S_{\V v}=\M{\Set P}$
if $\M{\Set P}^{\star}=\M{\Set P}$ and $\M{\Set P}^{2}=\M{\Set P}$.
Both of these conditions are met by the MAC discrete projection operator
$\M{\Set P}=\M I-\M D^{\star}\left(\M D\M D^{\star}\right)^{-1}\M D$,
which shows that our spatio-temporal discretization gives velocity
fluctuations that have the correct covariance (\ref{eq:C_vv_nodt}).
A straightforward extension of the analysis in Appendix \ref{sec:CNAccuracy}
shows that the Crank-Nicolson temporal discretization (\ref{eq:split_v})
gives the correct equilibrium velocity covariance for \emph{any} timestep
size, not just for small time steps. Further details will be presented
in future publications \cite{DFDB}.

\end{appendix}


\end{document}